\documentclass[aps,prb,showpacs,twocolumn,preprintnumbers,amsmath,amssymb]{revtex4}
\usepackage{dcolumn}% Align table columns on decimal point
\usepackage{bm}% bold math
\usepackage{amsmath}
\usepackage{feynmp}

\usepackage{graphicx}% Include figure files
\begin{document}

%\usepackage{graphicx}% Include figure files
%\usepackage{listings}
%\usepackage{amssymb,amsmath}
%\usepackage{color}
%\usepackage{wrapfig}
%\usepackage{caption}
%\usepackage{subcaption}
%\usepackage{geometry}
%\usepackage[toc,page]{appendix}

%\usepackage{hyperref}

%\lstset{language=matlab}

\title{Topological  Effects on the Magnetoconductivity in Topological Insulators}

\author{Vincent E. Sacksteder IV$^{1}$}
\email{vincent@sacksteder.com}
\thanks{Vincent Sacksteder previously worked at the Institute of Physics, Chinese Academy of Sciences, Beijing 100190, China, where this work was begun.}
\affiliation{Division of Physics and Applied Physics, Nanyang Technological University, 21 Nanyang Link, Singapore 637371}
%\affiliation{Institute of Physics, Chinese Academy of Sciences, Beijing 100190, China}

\author{ Kristin Bjorg Arnardottir }
\affiliation{Division of Physics and Applied Physics, Nanyang Technological University, 21 Nanyang Link, Singapore 637371}

\author{Stefan Kettemann}
\affiliation{Division of Advanced Materials Science, Pohang University of Science and Technology (POSTECH), Pohang 790-784, South Korea}
\affiliation{School of Engineering and Science, Jacobs University Bremen, Bremen 28759, Germany}

\author{Ivan A. Shelykh}
\affiliation{Division of Physics and Applied Physics, Nanyang Technological University, 21 Nanyang Link, Singapore 637371}
\affiliation{Science Institute, University of Iceland, Dunhagi-3, IS-107, Reykjavik, Iceland}

 \pacs{73.20.Fz, 73.43.Qt ,  73.23.-b, 73.25.+i  }

\date{\today}

\begin{abstract}
Three-dimensional strong topological insulators (TIs) guarantee the existence of a 2-D conducting surface state which completely covers the surface of the TI.  The TI surface state necessarily wraps around the TI's top, bottom, and two sidewalls, and is therefore topologically distinct from ordinary 2-D electron gases  (2DEGs) which are planar.  This has several consequences for  the magnetoconductivity $\Delta \sigma$, a frequently studied measure of weak antilocalization which is sensitive to the quantum coherence time $\tau_\phi$ and to temperature.  We show that conduction on the TI sidewalls systematically reduces $\Delta \sigma$,  multiplying it by a factor which is always less than one and decreases in thicker samples.  In addition, we  present both an analytical formula and numerical results for the tilted-field magnetoconductivity  which has been measured in several experiments. 
Lastly, we predict that as the temperature is reduced  $\Delta \sigma$ will enter a wrapped regime where it is sensitive to diffusion processes which make one or more circuits around the TI.  In this wrapped regime the magnetoconductivity's dependence on temperature, typically $1/T^2$ in 2DEGs,  disappears.    We present numerical and analytical predictions for the wrapped regime at both small and large field strengths.  The wrapped regime and topological signatures discussed here should be visible in the  same samples and at the same temperatures where the Altshuler-Aronov-Spivak (AAS) effect has already been observed, when the measurements are repeated with the magnetic field pointed perpendicularly to the TI's top face.  

%In addition, we predict  a new signature of the topological state: at low temperatures the magnetoresistance will deviate strongly from the Hikami-Larkin-Nagaoka (HLN) formula.  In this regime the magnetoresistance is dominated by  scattering processes which wrap around the TI sample.  The HLN formula's shoulder will  be replaced by a feature with a larger critical field magnetic strength that is caused by with wrapping.  Inside the wrapping regime the magnetoconductance will lose its dependence on temperature.  This new topological signature should be visible in the same samples and temperatures where the Altshuler-Aronov-Spivak (AAS) effect has already been observed.

\end{abstract}

\maketitle

\section{Introduction}

	       \begin{figure}[]
%\centering
\includegraphics[width=10cm,clip,angle=0]{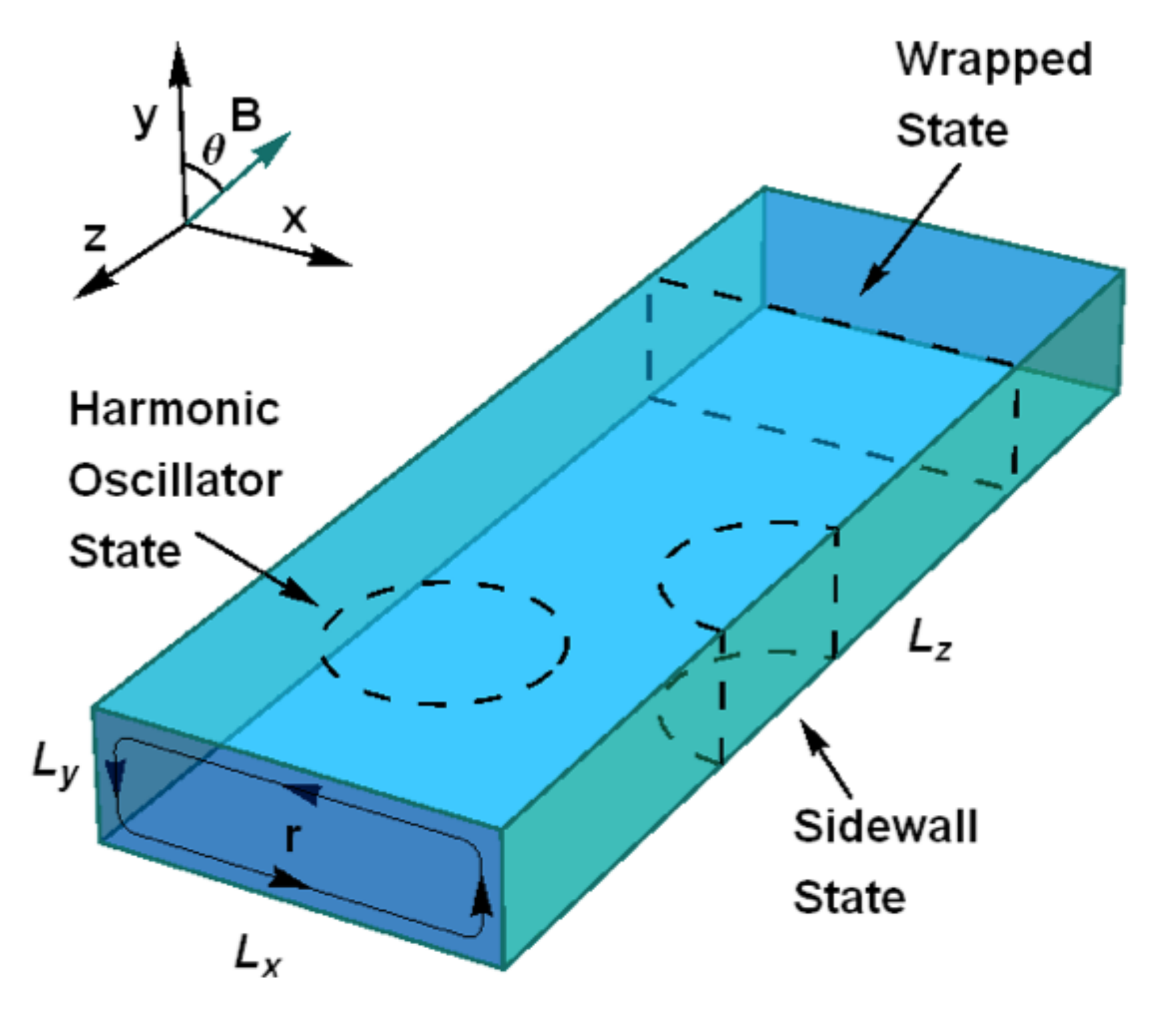}
\caption{ (Color online.)  The TI sample geometry and typical Weak Antilocalization states.  This article studies a TI sample of width $L_x$ and height $L_y$.  Conduction is along the $z$ axis, and the magnetic field $B$ is in the $x-y$ plane. "Perpendicular" fields are parallel to the $y$ axis and  perpendicular to the sample's top face, with $\theta = 0$.  "In-plane" fields  are parallel to the $x$ axis, with $\theta = \pi / 2$.  We use an $r$ coordinate which wraps the TI sample's perimeter, as shown by the arrowed line making a circuit around the front face.  The dashed lines illustrate three kinds of weak antilocalization states which contribute to the magnetoconductivity.  The Harmonic Oscillator WAL states execute cyclotron motion on a single face, the Sidewall states execute plane wave motion on the sidewalls and reflect off the top and bottom faces, and the Wrapped states make circuits around the entire TI perimeter. }
\label{Fig1Geometry}
\end{figure}

%Synchronize with HLN.  Their $a = 4x/\tau$ and they quote results in terms of $a \tau_i$ where the various $\tau_i$ are the scattering times related to different kinds of disorder.  The dephasing time is $\tau_\epsilon = -\imath \omega$.  

%\item While, Wentzel, Kramers, and Brillouin developed this approach in 1926, earlier in 1923, a mathematician, Harold Jeffreys, had already developed a more general method of approximating linear, second-order differential equations (the Schršdinger equation is a linear second order differential equation). Jeffreys is rarely given his proper credit.  JWKB method.  Jeffreys, Harold (1924). "On certain approximate solutions of linear differential equations of the second order". Proceedings of the London Mathematical Society 23: 428Ð436. doi:10.1112/plms/s2-23.1.428.  Maybe I don't need to cite them I just need to say JWKB.

The hallmark of topological insulators (TIs)  is metallic transport on their surfaces, coexisting with and reinforced by an insulating bulk.    \cite{Kane05, Min06, Zhang09, Hasan10, Culcer12, Bardarson13}  In diffusive samples where the scattering length is smaller than the sample size, the TI  surface state's conductivity is predicted to increase in the presence of scattering due a  quantum interference phenomenon called weak antilocalization (WAL). \cite{Ostrovsky07,Nomura07,Mucciolo10, wu2013effects}   The weak antilocalization is destroyed if time reversal symmetry is broken, so a small perpendicular magnetic field is sufficient to cause a clear drop in a TI's conductivity.   \cite{Chen10, Li12}  This negative magnetoconductivity at small fields is both the most accessible and the most frequently measured signal of weak antilocalization in TIs.    The  magnetoconductivity profile has a universal shape described in Hikami, Larkin, and Nagaoka's (HLN) celebrated paper, and frequently observation of this signal is used as a diagnostic determining the presence of 2-D surface transport. \cite{Hikami80}   
  
The HLN formula was derived for a 2-D electron gas (2DEG) moving on an infinite plane.  In an infinitely extended TI slab there would be two 2DEG layers, one on the TI's top and the other on the bottom.  If the top and bottom are decoupled then both the TI's conductance $G$ and its conductivity $\sigma = G L / W$ will simply be twice the HLN result, while if they are tightly coupled the doubling will be absent. ($L$ and $W$ are the sample length and width.)     This simple prescription for doubling the HLN magnetoconductivity has been confirmed by experiments on TI films which varied the coupling between the top and bottom surfaces. \cite{chen2011tunable, steinberg2011electrically, garate2012weak} 

This prediction suffers, however, from ignoring the structure of the topological insulator's surface state.  Any strong TI sample is completely enfolded by its surface state, as guaranteed by  topological protection.  In particular, the surface state occupies not only the TI's top and bottom, but also its sidewalls.  All of the TI's surfaces are coupled to each other, in contradiction to the simplistic HLN picture. Therefore  the finite TI sample size cannot be pushed to infinity, and  must  have an important role in the magnetoconductivity.  The current paper is devoted to a detailed investigation of the corresponding effects.

We will will start from consideration of a  TI surface state that wraps the entire sample and we will compute the perpendicular magnetoconductivity including the top, bottom, and sidewalls.  We will show that the sidewalls always cause a reduction of the conductivity with respect to the prediction of the HLN formula, because a perpendicular magnetic field does not destroy quantum interference processes which occur on the sidewalls.    This result will modify the interpretation of experimental measurements of the magnetoconductivity signal's magnitude, which is used as a diagnostic determining whether a TI sample's surfaces are strongly coupled.

A more intriguing result  occurs when quantum phase coherence survives long enough to allow a diffusing charge to make a circuit around the sample, i.e. when the diffusive coherence length $L_\phi$ is comparable to the sample perimeter $L_r$.  In this regime the magnetoconductivity is dominated by quantum interference between the paths making  one or more turns around the sample, and the HLN formula must be completely replaced.  In contrast with the HLN formula's prediction that the magnetoconductivity is governed by  $L_\phi$, in this wrapped regime the TI magnetoconductivity at low fields is controlled by  the sample width $L_x$.  The most distinctive signature of this effect is that $\sigma(B)$ will lose its temperature dependence, differently from  the HLN formula which scales with  $T^{-2}$.   The transition from wrapped to HLN behavior occurs at a characteristic field strength  $eB_w/\hbar = (4L_x l)^{-1}$  where $l$ is the scattering length.  It is remarkable that the magnetoconductivity, a result of diffusive physics at long length scales, is sensitive to the small scattering length scale $l$.  This is a consequence of equipartition of kinetic and potential energy in each Landau level state, as we will discuss later.  The $L_r \leq L_\phi$ regime discussed here has already been achieved in those experiments which have used a longitudinal magnetic field to observe Altshuler-Aronov-Spivak (AAS) oscillations;  all that is needed is to orient the field perpendicularly to the sample and repeat the magnetoconductivity measurement.  \cite{aas81,aronov1987magnetic}  AAS oscillations have been reported at $T = 2\;K$ in samples with perimeters $L_r$ as large as $500\; nm$, indicating that in those samples $L_\phi$ reached or exceeded the same value. \cite{peng2010aharonov,xiu2011manipulating,sulaev2013experimental}  Another experiment reported that $L_\phi$ exceeded $2$ microns at $T = 60 \; mK$. \cite{dufouleur2013quasiballistic}
% .... the perimeter is  350nm in peng, 500 nm and 2K in Sulaev, 430 nm and 1.4 K in Xiu11

  We will also report on the case when the magnetic field is oriented in-plane so that it pierces the TI sidewalls rather than the top and bottom.  All previous theory about the in-plane magnetoconductivity treated quantum wells where the charge carriers move freely between the top and bottom surfaces, or attributed it to the TI surface state's penetration into or through the bulk. \cite{Altshuler81,dugaev1984magnetoresistance,beenakker1988boundary,tkachov2013spin,raichev2000weak}  In-plane  signals in TIs sometimes have been attributed to the bulk contribution to conduction in the TI sample.  \cite{He11,Zhao13} Here we show that the TI sidewalls are  also a natural explanation of in-plane observations in TIs. In this case the sample height $L_y$, not the penetration depth $\lambda$, controls the in-plane effect, consistently with at least one recent experiment.  \cite{Lin13}

    Lastly we report analytical and numerical results on tilted fields  (i.e.  when the field orientation makes an arbitrary angle $\theta$ between the perpendicular and in-plane directions.) The corresponding  measurements have been used as diagnostics to confirm that the observed magnetoconductivity is sensitive only to $B \cos \theta$, not $B$  and $\cos \theta$ separately, as expected for a 2-D electron gas. \cite{chen2011tunable,He11, Wang12,Lee12,Cha12,Zhao13,Gehring13,Chen14} Most experiments find good agreement with the 2-D expectation only at small fields and small angles. \cite{He11,Cha12,Zhao13,Chen14} Some experiments have attributed the in-plane signal to the bulk and simply subtracted it from their results at other angles\cite{He11,Zhao13}, while one detailed experiment showed that this procedure could not match observations. \cite{Wang12}  Here we present  graphs of the magnetoconductivity corresponding to the tilted field results that experimentalists might expect to obtain and propose an approximate interpolating formula.

 The  structure of this article is as follows. In Section \ref{2DEGReview} we will review  the HLN formula for perpendicular fields and several results for in-plane fields, because these results are widely used and will be our reference points.  Next section \ref{Model} introduces our model, explains our method for evaluating the magnetoconductivity numerically, and details the important parameters and length scales.  Section \ref{Results} presents and discusses our numerical results, and we summarize our conclusions in section \ref{Conclusions}.  The Appendix presents certain analytical results on the magnetoconductivity.  
  
\section{\label{2DEGReview} Overview of the Results for 2-D Electron Gases}
For magnetic fields perpendicular to a 2-D electron gas, the Hikami-Larkin-Nagaoka  (HLN) formula has proved to give an adequate description of numerous experimental results.  The TI surface is a spin-1/2 system with strong spin orbit coupling, and spin polarization decays very quickly so only  the spin singlet channel participates in diffusive conduction.  \cite{sacksteder2012spin} In this case  the HLN formula reads:
\begin{eqnarray} \label{HLNFormula}
 \sigma_{HLN}(x) &=& \sigma(B=0)
\\ \nonumber 
 & + &\alpha \frac{G_0}{2 \pi}(\psi(1/2 + E_l/x) - \psi(1/2 + 1/x))
 \end{eqnarray}
where $G_0 = 2e^2 / h$ is the conductance quantum, $\psi(x)$ is the digamma function, and $\alpha = 1/2$ sets the magnitude.   $L_B = \sqrt{\hbar / 2 eB} $ is the magnetic length controlling quantum interference in a magnetic field, and the parameter $x = 2 L_\phi^2 / L_B^2 $ is proportional to the magnetic field strength $B$.  $L_\phi = \sqrt{D \tau_\phi} $ is the average distance that an electron covers before dephasing kills the corresponding quantum interference diagrams,  and $D$ is the diffusion constant.  $\tau_\phi$ is the dephasing time, which typically scales with $T^{-1}$ in 2-D diffusive systems and is sensitive to the dimensionality. The short-wavelength, high-energy cutoff on quantum interference processes is given by $E_l   = L_\phi^2/ l^2$,  where $l$ is the scattering length.  

At small $B$ (i.e. $L_B \gg L_\phi$) the HLN formula is a quadratic $\sigma_{HLN} \approx -\alpha \frac{G_0}{2 \pi} \frac{x^2}{24}$. It displays a shoulder at $L_B \propto L_\phi$ and transitions to a logarithmically decreasing curve at high fields.  At small fields $\sigma_{HLN}$ is typically proportional to $T^{-2}$, at large fields it scales with $\ln T$, and the shoulder occurs at  a value  of $B$ which typically scales with $T$.

Experimental fits to the HLN formula focus on three parameters:
\begin{itemize}
\item  Many experiments focus on $\sigma(B)$ at small fields where it is quadratic.  They measure the coefficient $b$ of the quadratic term, i.e. the coefficient $b$ of $\sigma_{HLN} =-  \alpha \frac{G_0}{2 \pi} \times b B^2$.  The HLN formula predicts that $b = {2 e^2 L_\phi^4}/{3 \hbar^2}$.
\item Using data at higher fields it is possible to measure the  diffusion length $L_\phi = \sqrt{D \tau_\phi}$, which determines the position of the HLN   shoulder.  Since only $\tau_\phi$ and not $D$ depends on temperature, magnetoconductivity measurements are a good probe of dephasing and its sources.
\item At higher fields one can also measure the magnitude $\alpha$.   If one plots $\sigma(B)$ as a function of $\ln B$, one should obtain a straight line for large $B$. The slope determines $\alpha$.  The HLN formula fixes $\alpha$ unambiguously at $\alpha = 1/2$  for a single 2DEG with strong spin-orbit coupling.  However experimentally $\alpha$ is observed to vary continuously in TI samples and to have values as large as $1$ (or even larger \cite{steinberg2011electrically}), and this has been attributed to the TI's  top and bottom surfaces.
\end{itemize}

Several authors have generalized the HLN formula to a pair of 2DEGs linked by random (disordered) jumps from one layer to the other, similarly to an infinite TI slab with no side surfaces.  \cite{bergmann1989weak, raichev2000weak,garate2012weak} This model adds an extra time scale $\tau_c$   characterizing the average time between interlayer jumps.  Its behavior  is strongly dependent on the ratio $\tau_c / \tau_\phi$ and on the coupling length scale $L_c = L_\phi \sqrt{\tau_c / \tau_\phi}$.   The exact result is the sum of two HLN curves $\sigma_a$ and $\sigma_b$.   In the weakly coupled $\tau_c / \tau_\phi \gg 1$ regime each of the two terms corresponds to the HLN result of one of the two 2DEGs; each 2DEG contributes independently and additively to the magnetoconductivity.  In the strongly coupled $\tau_c / \tau_\phi \ll 1$ regime the two terms $\sigma_a, \sigma_b$ have two distinct values of $L_\phi$ which depend strongly on the both $\tau_c$ and $\tau_\phi$.   The two $L_\phi$'s differ considerably.  At small fields $L_B \gg L_c$ either $\sigma_a$ or $\sigma_b$ dominates the magnetoconductivity, similarly to a single 2DEG.  Large enough fields disrupt the hopping between the 2DEGs and restore the magnetoconductivity to a sum of terms from two independent surfaces.  Practically speaking, these results indicate that at small fields the magnetoconductivity can be fit well with an HLN curve with variable magnitude $\alpha$.  At  strong coupling and weak fields one will find $\alpha = 1/2$, while at either weak coupling or strong fields $\alpha = 1$.

If we change the magnetic field's orientation to lie in-plane with the sample (and still perpendicular to the conduction), we obtain the problem of the in-plane magnetoconductivity.  This signal must be absolutely null in a 2DEG with zero depth because there is no magnetic flux through the 2-D plane.  Assuming that the electrons are free to diffuse throughout the vertical extent of the quantum well, Altshuler and Aronov published the first result which takes into account the 2DEG's depth.  \cite{Altshuler81} They studied a dirty film where the scattering length $l$ is much shorter than the sample thickness $L_y \gg l$ and obtained a logarithmic form:
\begin{eqnarray}
\sigma_{AA} &=& -\alpha \frac{G_0}{2 \pi}\ln |1 + b B^2|, \;
b  = e^2 L_y^2 L_\phi^2  / 12 \hbar^2 
\end{eqnarray}
Like the HLN result for perpendicular fields, the A-A formula for in-plane fields is quadratic at small fields and logarithmic at large fields, and displays a shoulder which transitions between the two.   Unlike HLN,  $\sigma_{AA}$ depends on the sample thickness, is  typically proportional to $L_\phi^2 \propto T^{-1}$ at small fields,  and its shoulder occurs at  a value  of $B$ which typically scales with $L_\phi^{-1/2} \propto T^{1/2}$.  Later  papers by Dugaev and Khmelnitsky and by Beenakker and Van Houten  showed that these scaling laws are general to any model where charge moves throughout the sample interior and the dephasing length $L_\phi$ is larger than the thickness.  \cite{dugaev1984magnetoresistance,beenakker1988boundary} Changing to a cleaner sample where $L_y \ll l$ only multiplies $b$ by $3 L_y / 8 l$, and changing the boundary conditions only multiplies $b$ by a numerical constant.

It is straightforward to apply the A-A approach to calculate the effect of the surface state's penetration into the bulk,  in  a half-infinite TI.  Tkachov and Hankiewicz reported that $b = 2 e^2 \lambda^2 L_\phi^2 / \hbar^2$ where $\lambda$ is the penetration depth. \cite{tkachov2013spin}

More interestingly, Raichev and Vasilopoulos (RV) calculated the in-plane magnetoconductivity of a pair of 2DEGs with random hopping.  \cite{raichev2000weak} This scenario is similar to a TI with no sides but with some random way for carriers to jump between the top and bottom surfaces.  At strong coupling $\tau_\phi / \tau_c \gg 1$ RV obtain an A-A type formula with $b = e^2 L_y^2 L_\phi^2 /4 \hbar^2$, but at strong enough fields $b B^2 \geq \tau_\phi / \tau_c$ the in-plane magnetoconductivity saturates and becomes constant.   At weak coupling $\tau_\phi / \tau_c \ll 1$ the magnetoconductivity is strongly suppressed, and in particular the quadratic coefficient $b$ at small fields is multiplied by $\frac{4}{3}(\tau_\phi / \tau_c)^2$.  

These results indicate that the coefficient $b$ measured at small  in-plane fields  is sensitive to the intersurface hopping $\tau_c$ and to the temperature $\tau_\phi \propto 1/T$, and this has been confirmed experimentally. \cite{Lin13}   They also indicate that $b$ is three times bigger when current moves only on a TI's top and bottom surfaces than its value when current moves in the bulk; $b$ can be used to measure the current distribution within a TI. \cite{MyInplane}    However these results give little insight into the physics to be expected from a TI's side surfaces, and no insight into topologically induced wrapping around the TI.  In particular, is the coupling from side surfaces a strong coupling, or a weak coupling?

\section{\label{Model}Weak Antilocalization  of a Wrapped Surface State}
%\textbf{ maybe maybe if time allows show a picture of a random walk returning to itself, with time-reversed paths, magnetic flux, etc.  Ivan would also like to have a diagram representing the Cooperon.}

% The focus of our study is conduction in the diffusive regime, after many scatterings.  In particular, we are not studying Aharonov-Bohm oscillations here, whixh occur only in the ballistic regime where the disorder-induced level smearing is small compared to the level spacing. 
 %  AB oscillations are a consequence  the finite spacing between energy levels that occurs in in a tube-like geometry - each energy level corresponds to a quantum of the angular momentum around the tube.

The focus of our study is weak antilocalization, a quantum interference process which either adds to or subtracts from the  probability that a charge will return to its starting point.   One speaks of weak localization or weak antilocalization depending on the sign of the quantum correction to the return probability.   On a TI surface  the return probability is decreased,  the conductivity increases, and this is called weak antilocalization.   This is caused by the strong spin-orbit coupling found in TIs.    In materials without  a strong spin-orbit coupling one obtains the opposite effect, a decreasing conductivity called weak localization.  The essential reason is that conduction is a long-distance phenomenon, and that the spin singlet channel and the three spin triplet channels  contribute with opposite signs to the quantum interference diagrams.  In materials with weak spin-orbit coupling both the spin singlet and the spin triplet persist over long distances, the three spin triplet channels win, and the net effect is an increased return probability.  In contrast, it has been established both theoretically and experimentally that in TIs the spin relaxation length is very short. \cite{sacksteder2012spin,chen2011tunable,Wang12}  Therefore only the spin singlet contributes to quantum interference, producing a decreased return probability and weak antilocalization.

The absence of long-distance spin physics also allows us to neglect  the Zeeman energy, which ordinarily would mix the triplet channel with the singlet channel and thus destroy the weak antilocalization signal  which is the focus of our article.  \cite{maekawa1981magnetoresistance}  It is also known that Coulomb interactions combined with the Zeeman energy cause additional changes to weak antilocalization.  \cite{lee1985disordered,altshuler1982spin}  Fortunately both theory and experiment also verify that these effects disappear when the spin decay time is very short, as is the case on a TI surface.  \cite{lee1985disordered,altshuler1982spin,sahnoune1992influence, chen2011tunable,Wang12}   In conclusion, the physics of weak antilocalization on the surface of a TI, because it concerns only long length scales, is entirely determined by diffusion of the spin singlet.  In particular, spin triplet diffusion does not need to be modeled, and  the effect of interactions on weak antilocalization can be neglected.    At the long distance scales which concern us the only net effect of the TI's spin-orbit coupling are to destroy the spin polarization and reverse the overall  sign of the quantum corrections.  Therefore we will model only the singlet channel of quantum interference  in a magnetic field. 
  %    \cite{Wang12} and also Yongqing's earlier article \cite{chen2011tunable} do a B cos plot and find great agreement  out to 1 T and conclude that spin-orbit really is killing the triplet and Zeeman+ interaction effects.  In the earlier article they also make the same point by studying how temperature changes the logarithmic conductivity.    In \cite{Wang12} they also plot the MC's variation at angles just off from in-plane,  and they also find that for their samples a recipe of subtraction the in-plane and then doing B cos theta won't account for things.  And they suggest the A-A formula.  \cite{chen2011tunable} cites four articles that say that spin-orbit kills Zeeman and interaction and triplet effects.

 In the diffusive regime, after many scatterings, the quantum correction to the conductivity at a position $\vec{x}$ is equal to:
\begin{eqnarray}
\sigma_{WAL}(B, \vec{x}) & = &   G_0  \frac{   D \tau}{ \Gamma^0}    \; \langle \vec{x} | \Gamma(B) | \vec{x} \rangle, \; 
\nonumber \\
 \Gamma(B) &= & \Gamma^0  (\tau / \tau_\phi- D {\tau}\,(\vec{\nabla} - \imath 2e \vec{A}/\hbar )^2)^{-1}
\end{eqnarray}
where $\Gamma^0 = (2 \pi \nu \tau)^{-1}$ describes the scattering strength and scatterer density, $\nu$ is the density of states,  $\tau$ is the scattering time,  and $\vec{A}$ is the gauge field associated with the magnetic field\footnote{If we had included the three spin triplet states in our calculation of $\sigma_{WAL}$, then $\Gamma(B)$ would be a $4 \times 4 $ operator acting on the space of spin singlet and spin triplet states, and $\sigma_{WAL}$ would be determined by $\Gamma$'s trace.   Because the spin triplet has short decay times, its contribution to $\Gamma(B)$'s trace is small.}. \cite{Hikami80, rammer2004quantum, mccann2006weak, garate2012weak}

The mathematical content of this formula is very simple - it simply calculates the return probability of a particle that starts at $\vec{x}$ and executes a random walk in a magnetic field $\vec{B} = \vec{\nabla} \times \vec{A}$.  The result of many scatterings is diffusion, which is described by the diffusion kernel $D {\tau}\,(\vec{\nabla} - \imath 2e \vec{A}/\hbar )^2$.  The additive constant $\tau / \tau_\phi$ describes gradual extinction of the random walker at the dephasing time scale $\tau_\phi$, and the matrix element  $ \langle \vec{x} | \Gamma(B) | \vec{x} \rangle$ extracts the random walker's return probability. A last detail is left implicit in our formula:  an ultraviolet cutoff at the scattering length $l$.  We are concerned with random walks with an average step length $l$, so we exclude length scales smaller than $l$ and energy scales higher than $E_l  =  L_\phi^2 / l^2$.

The only tricky issue here is the  identity of the random walker.  It is  not electronic charge, as can be seen from the factor of $2e$ - not $e$ - which multiplies the gauge potential.  Instead it is the Cooperon, a disorder-induced correlation between the electron wave-function $\psi$ and  its complex conjugate $\psi^\dagger$,  which is completely responsible for both weak localization and weak antilocalization.  The correlation in question here is of a very special nature: the Cooperon describes correlations where on one hand  $\psi$ executes random walks, while on the other hand $\psi^\dagger$ executes time-reversed copies of the same random  walks.  For this reason the Cooperon can not be reduced to classical physics; it is a distinctly quantum phenomenon.  For the same reason the Cooperon is extremely sensitive  to anything which breaks time reversal symmetry, which is why weak antilocalization has such a distinctive signal at small magnetic fields.

 $\sigma_{WAL}(B, \vec{x}) $ is an inherently position-sensitive quantity, describing the quantum probability of returning to point $\vec{x}$.  It is sensitive to the   device geometry and also to spatial variations in the diffusion constant, dephasing time, and magnetic field. In previous works on quantum wells such spatial dependence has been neglected because the quantum well's geometry is well approximated by an infinite plane with translational invariance.  Here we consider TIs, where the conduction occurs  on a surface which  always completely wraps the TI bulk.  Therefore we must track explicitly spatial variations in  $\sigma_{WAL}(B, \vec{x}) $.  The dominant source of these variations is changes in the amount of magnetic flux piercing the surface.  We will neglect subleading effects, such as the sensitivity of the surface state dispersion to the crystallographic orientation of the TI surface. \cite{zhang2012surface, silvestrov2012spin} Variations in the diffusion constant and dephasing time, whether within the TI's surfaces or when crossing the edges between surfaces, are outside the scope of this article, and we will fix these constants to spatially uniform values.  
 
 %The point of this article is that a TI's magnetoconductance cannot be modeled accurately  with a translationally invariant plane, and that wrapping-around of the surface conduction is crucial to a correct computation.

 The experimental observable that has been measured widely is not the local conductivity  $ \sigma_{WAL}(B,\vec{x})$, but instead the global conductivity  $\sigma = G L / W$ of the TI device.   In the diffusive regime where electrons diffuse freely around the TI surface, the global conductivity is equal to the spatial average of $\sigma_{WAL}(B,\vec{x})$, which will be our focus:
 \begin{eqnarray}
\langle \sigma_{WAL}(B)\rangle & = &   G_0    L_\phi^2  \; \int \frac{d^2\vec{x}}{L_r L_z}   \; \langle \vec{x} | \tilde{\Gamma}(B) | \vec{x} \rangle, \; 
\nonumber \\
 \tilde{\Gamma}(B) &= &   (-L_\phi^2 \,(\vec{\nabla} - \imath 2e \vec{A}/\hbar )^2 + 1)^{-1}
\end{eqnarray}
 The ${d^2\vec{x}}$ integral covers the entire surface of the TI sample. $L_r$ is the sample's perimeter  and $L_z$ is its length, so $L_r L_z$ is the surface area.

We note immediately that because  Cooperon kernel is diffusive, and because a decay term equal to $1$ is added to $-L_\phi^2 \vec{\nabla}^2$,    the Cooperon decays exponentially at the dephasing length $L_\phi$.  Therefore  weak antilocalization is not sensitive to the TI surface's wrapping around the sample, unless the perimeter $L_r$ is either comparable to or smaller than $L_\phi$.   This same condition  regulates the existence of Altshuler-Aronov-Spivak (AAS) oscillations, another weak antilocalization signal which occurs  when the magnetic field is oriented longitudinally with the axis of conduction.

\subsection{Reduction to a One Dimensional Problem}
  In a magnetic field a charged particle executes circular motion around a slowly moving point called the guiding center that drifts in response to external forces.  In our problem there are no forces other than the magnetic force, so the guiding center's momentum is conserved.   We take advantage of this and choose a gauge potential which preserves translational invariance along the $\hat{z}$ axis of conduction.  This technique effectively solves the guiding center's motion. 
We choose a gauge potential $\vec{A} = A_z \hat{z}$ that is aligned with the $\hat{z}$ axis.  We restrict  it to depend only on $r$, the coordinate which wraps around the TI surface.  These choices simplify the Laplacian to  $(\vec{\nabla} - \imath 2e \vec{A}/\hbar )^2 =   -(k_z - 2 e A_z(r) /   \hbar)^2 +  (\partial/\partial_r)^2 $.   We have preserved translational invariance along the $\hat{z}$ axis and therefore can  resolve the Cooperon with the $z$ momentum $k_z$.  A simple exercise obtains the equations which we will use for our numerical calculations:
\begin{eqnarray} \label{MCequation}
\langle \sigma_{WAL}(B) \rangle & = &   \frac{G_0 }{4\pi}   \;x \; \int \frac{dR_g}{L_r} \; \int {dr}   \; 
 \nonumber \\
 & \, & 
 \langle r | ( V(r, R_g) - L_\phi^2 (\partial/\partial_r)^2  + 1)^{-1}| r \rangle, \; 
\nonumber \\
 & = &  \frac{G_0 }{4\pi}   \;x \; \int \frac{dR_g}{L_r} \;  {Tr} (\,A^{-1}\,)
 \nonumber \\
 & = &  \frac{G_0 }{4\pi}   \;x \; \int \frac{dR_g}{L_r} \; \sum_{i}^{E_i < E_l} E_i^{-1}, \; E_i | i \rangle =  A | i \rangle
 \nonumber \\
 A = & \;&   V(r, R_g) - L_\phi^2 (\partial/\partial_r)^2  + 1, \; 
\nonumber \\
V(r, R_g) &=&  ( x^2 / 4 L_\phi^2) (R(r) - R_g)^2
\nonumber \\
R(r) &=& {- 2 e A_z(r) L_B^2 /\hbar},\, R_g = -L_B^2 k_z
\end{eqnarray}
Here and elsewhere $x = 2 L_\phi^2 / L_B^2$ is a dimensionless variable that is proportional to the magnetic field strength $B$. We now have an integral over the guiding center's longitudinal momentum $k_z$, which is linearly related to $R_g = -L_B^2 k_z$, the guiding center's equilibrium position on the $r$ coordinate which parameterizes the  TI's perimeter.  For each particular value of $R_g \propto k_z$ we must solve a 1-D Schrodinger equation $E_i | i \rangle =  A | i \rangle = (V(r, R_g) - L_\phi^2 (\partial/\partial_r)^2  + 1 ) | i \rangle$ describing the problem of cyclotron motion in the TI's magnetic field.   

In the limit of an infinitely wide 2-D surface with no boundaries or sidewalls, the potential $V(r,R_g)$ in equation \ref{MCequation} is quadratic, and the eigenvalue problem  $E_i | i \rangle =  A | i \rangle$ maps exactly to the Landau level problem.  The  energies are $E_i = ( i + 1/2 )  \omega_C + 1 $ where  $ \omega_C = 2 L_{\phi}^2/L_B^2 = x$ is the cyclotron frequency of the Cooperon.  Equation \ref{MCequation}'s sum over the equally spaced Landau level energies $E_i$ can be summed analytically, producing the HLN formula found in equation \ref{HLNFormula}.

In a TI with top and bottom surfaces and sidewalls the potential $V(r,R_g)$ is not a simple quadratic, which will force us to use numerical not analytical methods to evaluation equation \ref{MCequation}. If the TI sample's height and width are $L_y, L_x$ respectively and the magnetic field is aligned at an angle $\theta$ with respect to the $y$ axis which lies normal to the TI's top and bottom surfaces, then the $R(r)$ function figuring in the potential is:
\begin{widetext}
 \begin{eqnarray}
R(r) &= & 
\begin{cases} (r-L_x/2) \cos \theta - (L_y / 2) \sin \theta, & 0 < r < L_x
\nonumber \\
 (L_x/2) \cos \theta + ( r -   L_x - L_y/2)  \sin \theta,  &L_x < r < L_x + L_y \; 
\nonumber \\
(-r+ 3 L_x/2 + L_y)  \cos \theta + (L_y/2) \sin \theta, & L_x + L_y < r <  2 L_x + L_y \;  
\nonumber \\
  - (L_x/2) \cos \theta + (-r +2 L_x + 3 L_y/2) \sin \theta,  & 2 L_x + L_y < r < 2 L_x + 2 L_y \; 
 \end{cases}
\end{eqnarray}
\end{widetext}
The second moment of $R(r)$ is:
\begin{eqnarray}\label{R2Formula}
  \langle R^2(r) \rangle &= & \frac{(L_x/3 + L_y) L_x^2 \cos^2\theta + (L_x + L_y/3) L_y^2 \sin^2 \theta }{4 (L_x + L_y)}
  \nonumber \\
 \end{eqnarray}

On each surface we have a quadratic potential $V(r,R_g) = ( x^2 /4 L_\phi^2)(R(r) - R_g)^2$ with minima located at $R_g$.  This potential describes the fact that  eigenstates  are constrained to orbit $R_g$.   A state with energy $E$ has cyclotron radius  $(2 L_\phi / x) \sqrt{E}$, and decays exponentially outside that radius.  The largest permitted cyclotron  radius is determined by the ultraviolet cutoff $E_l = L_\phi^2 / l^2$, is equal to $\delta R = L_B^2 /l = {2 L_\phi^2}/{ l x}$, and scales with $ B^{-1}$. 

Since $R(r)$'s range is limited to the interval $\left[-(L_x |\cos \theta| + L_y |\sin \theta|)/2, +(L_x |\cos \theta| + L_y |\sin \theta|)/2 \right]$, we can  set $R_g$'s limits of integration at $\pm((L_x |\cos \theta |+ L_y |\sin \theta|)/2 + \delta R)$.

\subsection{Classification of Cooperon Eigenstates}
   \begin{figure}[]
%\centering
\includegraphics[width=9cm,clip,angle=0]{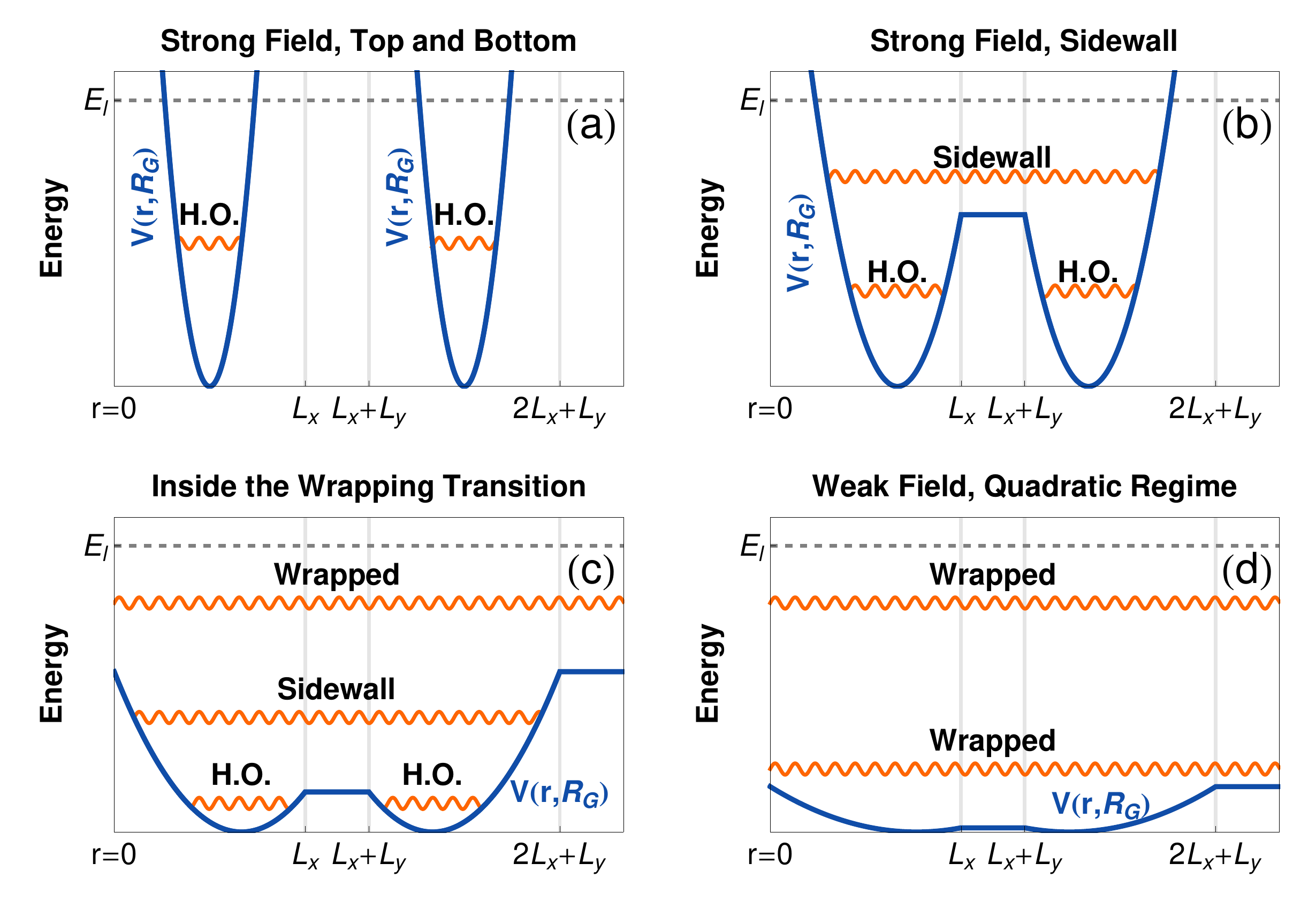}
\caption{ (Color online.) The potential and the Cooperon eigenstates.   The grey dashed lines sketch the potential $V(r,R_g)$ as a function of the coordinate $r$ which wraps around the TI sample.  $E_l = L_\phi^2 / l^2$ represents the ultraviolet cutoff.  The orange wavy lines illustrate typical harmonic oscillator, sidewall, and wrapped states.  Pane (a) illustrates a strong magnetic field with $R_g = L_x/2$ positioned in the middle of the top face; all states are harmonic oscillator (H.O.) states centered at $R_g$.  Pane (b) illustrates a strong magnetic field with $R_g = 0.8 L_x$ near the sidewall, allowing some sidewall states.  Pane (c) illustrates a weaker field inside the wrapping transition, allowing wrapped states. Pane (d) illustrates a very weak field where all states wrap the TI. }
\label{Fig1Potential}
\end{figure}

Equation \ref{MCequation}'s  structure, which is centered on a Schrodinger equation $E_i | i \rangle =  A | i \rangle$ with a non-uniform potential,  indicates that the eigenstates $| i \rangle $ are crucial to determining the  magnetoconductivity.  These are not electron states.  They are eigenstates of the Cooperon operator,  the diffusive operator  which controls the  interference diagrams that cause weak antilocalization.  We can  immediately distinguish four types of  Cooperon eigenstates by comparing their eigenvalues $E_i$ to the maximum value of the potential, which is $V_{max}(k_z) = L_\phi^2  ((L_x/2 L_B^2) |\cos \theta|  + (L_y/2 L_B^2) |\sin \theta| +  |k_z|)^2  $.  The four types of states are:
  \begin{enumerate}
  \item \textbf{Wrapped States.}   If the energy $E_i$ is large enough that it exceeds  $V_{max}$, then the state wraps around the entire TI nanowire.
 \item \textbf{Harmonic Oscillator states.}  If the energy $E_i$ is smaller than  $V_{max}$, AND if  $E = E_i$ intersects the potential at two points on  \textit{the same face} of the TI, then the state is trapped on that face and executes cyclotron motion there.    
 \item \textbf{ Sidewall states.}  If the  energy is smaller than  $V_{max}$, AND  if  $E = E_i$ intersects the potential on \textit{two opposite faces} of the TI (for example the top and bottom), then the state is spread across three faces of the TI.  We focus on the particular case where the magnetic field $B$ is perpendicular to the TI top surface; in this case the state moves as a plane wave on one of the TI's sidewalls and penetrates only a small distance into the top and bottom faces, where it executes cyclotron motion.  We call this a sidewall state.
  \item \textbf{ Edge states.}  If the  energy is smaller than  $V_{max}$, AND  if  $E = E_i$ intersects the potential on \textit{two adjacent faces} of the TI, then the state is trapped at the  corner between those two faces; it is an edge state.  This type of state occurs only when the magnetic field is at an angle relative to perpendicular, i.e. when $\theta \neq 0$.  
 \end{enumerate}

\subsection{Numerical Implementation}
All of our numerical results in the next section were obtained by numerically evaluating  equation \ref{MCequation}.  Our numerical implementation is straightforward.  We evaluate $- L_\phi^2 (\partial/\partial_r)^2 $ by discretizing on a lattice in position space with lattice spacing $a$; we obtain $- L_\phi^2 (\partial/\partial_r)^2  = 4L_\phi^2 \sin^2(k_r a/2) /a^2$.  We choose $a = l \sqrt{2}$, so the maximum value of the kinetic term is $2 E_l  $. The finite lattice spacing does introduce certain numerical errors, but these have no physical significance because they just slightly redefine the high-momentum cutoff at $E_l  = L_\phi^2 / l^2$.

To obtain a numerically efficient calculation  we cap the potential at $ 2 E_l $, so that the spectrum is bounded above at about $4 E_l$.  For safety we take the upper bound as $4.1\; E_l$.

Numerically we implement the ultraviolet $E_l = L_\phi^2 / l^2$ cutoff using this formula:
\begin{eqnarray}
\nonumber
{Tr} (\,A^{-1}\,) & = & {Tr} (\mathcal{D}/A), \; \mathcal{D} = \frac{1}{2}(1 - \tanh((A - E_l) / \delta))
\end{eqnarray}
%& \approx & 4x \sum_{n=0}^{E_l / 4x} (4 x (n+1/2) + 1)^{-1} = \sum_{n=0}^{E_l / 4x} ( n+1/2 + 1/4x)^{-1} 
%\nonumber \\
%&=&  dg(3/2 + (E_l + 1)/4x) - dg(1/2 + 1/4x)
The operator $\mathcal{D}$ is just a smoothed step function.  $\delta$ sets the  width of the energy cutoff, which is centered around $E_l = L_\phi^2 / l^2$.  The $\tanh$ function inside of $\mathcal{D}$ is approximated by a Chebyshev polynomial, and we choose $\delta$ to match the polynomial's order $K$;  $\delta \approx 4 E_l/ K$.     We have checked that all of our results have converged with respect to $K$.
%$\delta = E_l / K$ and $K$ are two parameters which set the width of the energy cutoff around $E_l = L_\phi^2 / l^2$.  When $l = 1/80$ and $W = 0.5$ I usually get convergence when $K = 150$, so $\delta = E_l / K \approx 40$. Maybe it's the basis size, $W/l=40$.  

We discretize the $R_g$ integral on a grid with  $2N_z + 1 $ points and check for convergence with respect to $N_z$.  Generally speaking, convergence occurs very quickly once $N_z$ reaches a threshold  value that is proportional to the sample perimeter $L_r$.  

\subsection{Important length scales and parameters}

Our model has the following parameters:
\begin{itemize}
\item $L_\phi = \sqrt{D \tau_\phi}$,  the dephasing length which regulates the Cooperon decay. This is the length that a particle can diffuse before dephasing occurs.  In diffusive 2-D systems $\tau_\phi$ scales inversely with temperature $T$, so $L_\phi \propto T^{-1/2}$.
 \item $L_x$, $L_y$, and $ L_r = 2 L_x + 2 L_y$, the sample width, height, and perimeter.  We will show that when these scales are long compared to $L_\phi$  the HLN result is, at leading order, simply reduced by a factor of $ 2 L_x/  L_r$.
 \item $l$, the scattering length, which determines the Cooperon's ultraviolet cutoff $E_l = L_\phi^2 / l^2$. 
\item $\theta$, the angle of the magnetic field.  
\item $B$, the magnetic field strength.  Instead of $B$ we will usually discuss the dimensionless  variable $x = 2 L_\phi^2 / L_B^2  = 4 e L_\phi^2 B / \hbar$, which is proportional to $B$.  Our model gives spurious results  when $B$ is strong enough that cyclotron radius is near the scattering length.  Practically speaking, this occurs when $x/E_l  =2 l^2/ L_B^2 >  \left[0.3, 1 \right]$.
\item $ \delta R = L_B^2 /  l = 2L_\phi^2 / l x$,  the cyclotron radius of the highest Landau levels at energies near  the ultraviolet cutoff $E_l$.
\end{itemize}

\section{\label{Results}Results}

\subsection{The geometric factor at large $B$. }

A TI sample's sidewalls are not pierced by a  magnetic flux when the magnetic field is perpendicular to the sample's top and bottom faces.  Therefore quantum interference diagrams occurring only on the sidewalls are completely unaffected by the magnetic field.  This reduces the sample's average magnetoconductivity, as compared to a simple TI model without sidewalls.

The magnetoconductivity including sidewalls is, in the most general case, difficult to compute because some interference diagrams visit both the sidewalls and the sample top and bottom, and this computation is the main focus of this paper.  However a simpler case exists which is relevant to most TI weak antilocalization experiments, where  the decoherence length $L_\phi$ is small compared to the sample height $L_y$ and width $L_x$.  In these experiments only a small fraction of the quantum interference involves two or more of the sample's faces, and the following result is easily derived: 
\begin{eqnarray} \label{GeometricFactorEq}
\langle \sigma_{WAL}(B) \rangle  &=&  \frac{2 L_x }{L_r} \sigma_{HLN}(x ) + O(L_\phi/L_x) + O(L_\phi/L_y)
\nonumber \\
% + 2 \langle \sigma_{s}(B) \rangle, \; 
 %\nonumber \\
%\langle \sigma_{s}(B) \rangle & = &  \frac{G_0 }{4\pi}   \;x \; \int_{-\delta R}^{\delta R} \frac{dR_g}{L_r} \; \int {dr}   \; \langle r | A^{-1}| r \rangle, \; 
%\nonumber \\
%& - & \frac{2 \, \delta R}{L_r}  \sigma_{HLN}(x)
 \end{eqnarray}
$\sigma_{HLN}(x )$ is the HLN conductivity of two separate TI surfaces.   The point is simple:  the HLN conductivity is multiplied by a factor of $2 L_x / L_r < 1$.    Figure 3 confirms this result by plotting the magnetoconductivity in samples of fixed width $L_x = 10$, as the sample height $L_y$ is varied from $30$ to $0$.  The magnetic field strength is kept fixed at four values $L_B^2 = \left[1.25, \, 2.5, \, 5, \, 10 \right] $, the dephasing length is $L_\phi = 5$, and the scattering length is $l = 0.5$.  When the sample is of height $L_y = 30$  we expect to obtain $0.25 \;\sigma_{HLN}(x)$, while in an  infinitely thin sample we expect the unmodified HLN result, and we  expect a linear dependence on   $2 L_x/L_r$ in  intermediate cases.

Figure 3 shows our numerical results, which were obtained by evaluating  equation \ref{MCequation}. They confirm Eq. \ref{GeometricFactorEq} when $2 L_x/ L_r < 0.5$; i.e. when the sample  thickness $L_y$ is large enough to exceed two decoherence lengths $L_y > 2 L_\phi$.  In  thinner samples  with $L_y < 2 L_\phi$ Figure 3 reveals that our linear formula overestimates the magnetoconductivity.   In other words, even in infinitely thin samples with no sidewalls the magnetoconductivity is always less than the HLN result.  We expect this undershoot in the thin limit to be proportional to $L_\phi / L_x$.

  \begin{figure}[]
%\centering
\includegraphics[width=8cm,clip,angle=0]{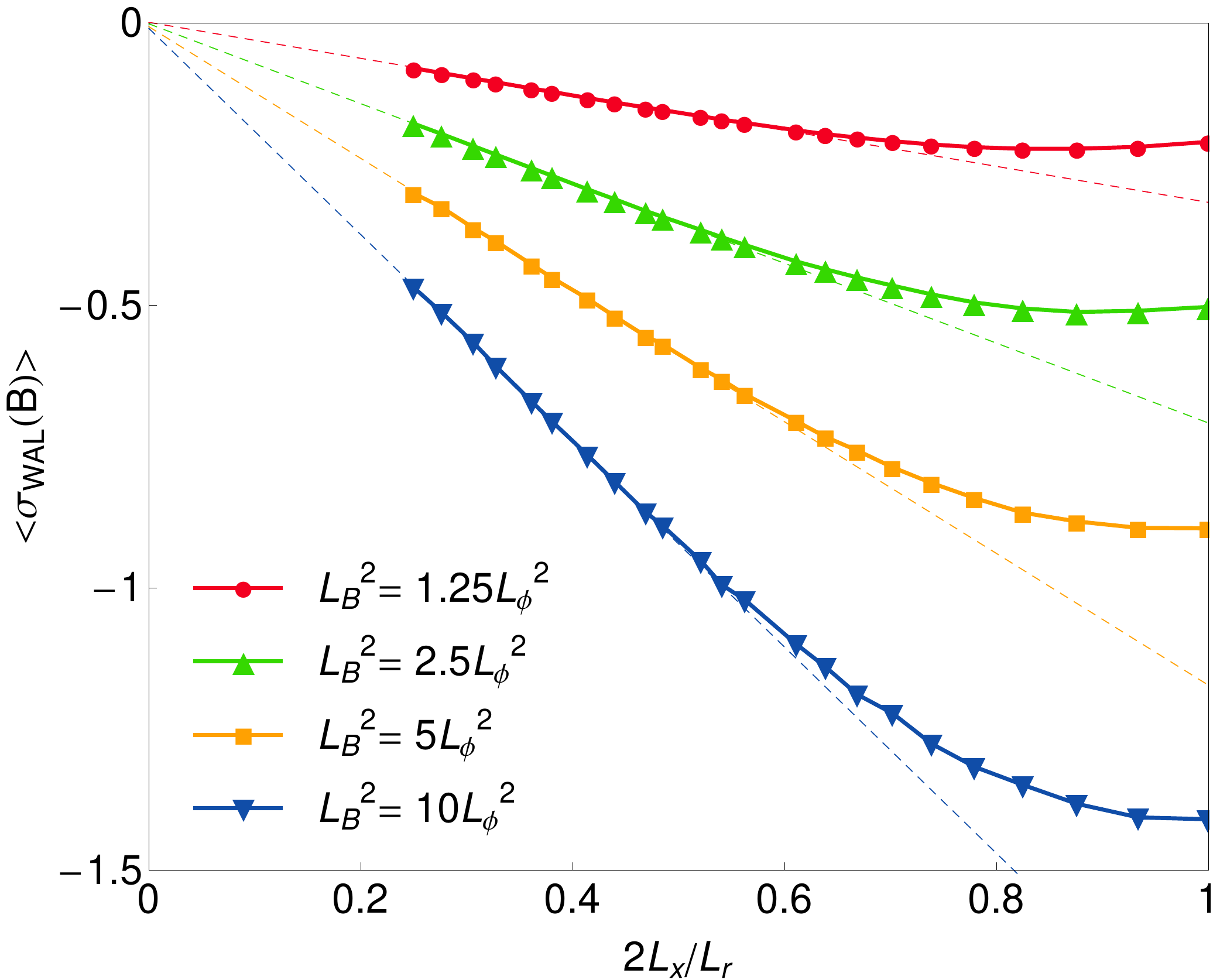}
\caption{ (Color online.)  The TI average magnetoconductivity $\langle \sigma_{WAL}(B) \rangle$ is reduced by the TI sidewalls. The solid lines plot $\langle \sigma_{WAL}(B) \rangle$ as a function $2 L_x / L_r$, where $L_x$ is the sample width and $L_r$ is the sample perimeter.  The four lines, from top to bottom, hold the magnetic  field  fixed at $L_B^2 = \left[1.25, 2.5, 5, 10 \right] L_\phi^2$.  The coherence length is  much smaller than the sample perimeter.   The dashed lines are linear fits to the data, and their excellent agreement with the data at $2 L_x / L_r < 0.5$ demonstrates that sidewalls multiply the HLN conductivity by $2 L_x / L_r $.   The deviation from linearity in thin samples shows that even in infinitely thin samples the conductivity is less than the HLN prediction because the top and bottom are coupled at the TI's sides. }
\label{FigAspectRatio}
\end{figure}

These results apply only when $L_\phi$ is small compared to the sample size.  We also have studied smaller samples, and confirmed that the linear scaling disappears.  

In conclusion, at small $L_\phi$ a TI with sidewalls will never attain the HLN prediction.  In thin samples the real conductivity will reduced by a term of order $L_\phi / L_x$, while in thicker samples the HLN result will be multiplied $2 L_x / L_r$.  This result will be of considerable interest to WAL experiments where the magnetoconductivity's magnitude has been used to count the number of independent conducting 2-D channels and to analyze the coupling strength between a TI's top and bottom surfaces.

\subsection{The Regime of Wrapped Cooperon  States }

We now turn toward systems  whose perimeter $L_r$  is comparable to or smaller than the dephasing length $L_\phi$, bringing weak antilocalization into a regime that is sensitive to each of the TI's faces.   Figure 4a shows our results for the magnetoconductivity $\langle \sigma_{WAL}(x) \rangle $ as a function of the dimensionless parameter $x$, which is proportional to $B$.  Here we treat the simple case of thin samples with height $L_y= 0$, and we will discuss thicker samples later. The scattering length is $l = L_\phi / 100$.  The black line on the left of Figure 4a is the HLN curve, which is realized in samples with width $L_x$ far exceeding $L_\phi$.  Moving to the right, the solid curves with symbols show successively samples with widths  $L_x = \left[0.8, 0.4, 0.2, 0.1 \right] L_\phi$.  The main qualitative features are immediately clear: in finite samples  the sensitivity to magnetic field is reduced, and the     HLN curve's shoulder at $x \approx 1$ is  replaced by a structure that moves steadily to higher magnetic fields as the sample size decreases.  Attempts to fit this magnetoconductivity to the HLN curve will obtain a value of $L_\phi$ which  is  reduced by a large multiplicative factor controlled by the sample width $L_x$.  The fitted value of $L_\phi$ will be proportional to $L_x$, not to the much larger value of $L_\phi$.

Key to these results is the radius of cyclotron orbits, because this length scale limits exploration of the TI perimeter.   The cyclotron radius is larger for higher excited states, and is $\delta R = L_B^2 / l$ at the ultraviolet cutoff $E_l = L_\phi^2 / l^2$.   At high fields $\delta R$ becomes so small that states are unable to cross between faces, and we obtain the modified HLN formula $\frac{2 L_x }{L_r} \sigma_{HLN}(x)$, which is proportional to $\ln x$.  Therefore  at high fields we always obtain the HLN logarithm.   This is confirmed by the high-field sector (right side) of Figure 4a.     In this semilog plot all five curves are linear at large $B$, matching nicely the straight lines.  This linear behavior on our semilog plot shows that the conductivity increases logarithmically with the system size, exactly as predicted by the HLN formula.

   \begin{figure}[]
%\centering
\includegraphics[width=9cm,clip,angle=0]{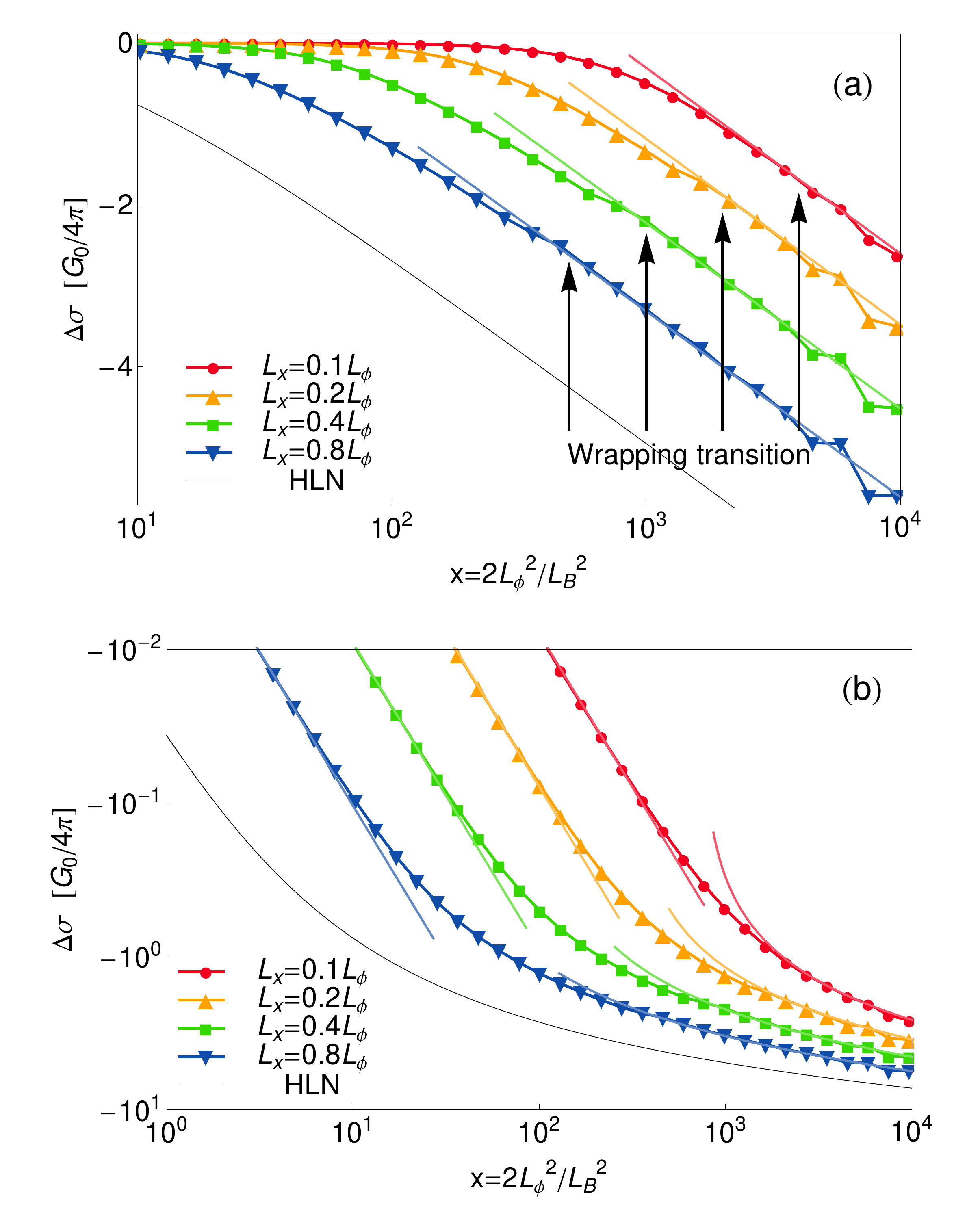}
\caption{ (Color online.)  The Wrapping Transition. Pane (a) shows the magnetoconductivity $\langle \sigma_{WAL}(B) \rangle$ as a function of $x \propto B \propto L_B^{-2}$.  The four lines are for four sample widths, from left to right $L_x = \left[0.8, 0.4, 0.2, 0.1 \right] L_\phi$.  The HLN curve is plotted in black.  At high fields the data matches logarithms similar to the HLN curve, seen as straight lines on this semilog plot.  At $x = 4 L_\phi^2 / L_x l$ the data departs from a logarithm because Cooperon eigenstates begin to wrap around the TI sample, becoming sensitive to its topology.  Pane (b) shows the same data on a log-log plot.  The small-field data matches well to  quadratic curves, seen here as straight lines, and is controlled entirely by wrapped states. The small-field regime ends at $L_B^2 < L_x^2/2$, when the potential becomes strong enough to cause bound states. }
\label{FigWrap}
\end{figure}

The HLN logarithm terminates at small fields where $L_B \propto L_\phi, \; x \propto 1$, and transitions into quadratic behavior.  Figure 4a shows that a TI sample begins to deviate from HLN  logarithmic behavior at far larger field strengths.  The deviation begins with a wrapping transition,  where the cyclotron radius becomes large enough to allow states to completely wrap the TI sample.    Comparison of the magnetic potential $V(r, R_g)$  with the ultraviolet cutoff  at $E_l = L_\phi^2/l^2$ indicates that the highest energy states start to  wrap the sample when $L_x |\cos \theta| + L_y |\sin \theta |= 2 \delta R $, which in a perpendicular field  simplifies to $L_B^2 = L_x l / 2$.    This is confirmed by our numerical results shown in Figure 4a.   The wrapping formula predicts  that the wrapping transition is at $x =  \left[ 500, 1000, 2000, 4000 \right] $ for the $W = \left[ 0.8, 0.4, 0.2 , 0.1 \right] L_\phi$ samples.  For the right-most line ($W = 0.1\, L_\phi$) the predicted transition value $x = 4000$ puts $L_B$ too close to the scattering length $l$ to obtain good accuracy, and the wrapping transition is hard to pin down.  However the three other lines at $W = \left[ 0.8, 0.4, 0.2 \right] L_\phi$   show clear wrapping transitions.  The arrows highlight the points where the numerical data show  clear wrapping transitions, and these points agree very well with our wrapping formula, $L_B^2 = L_x l / 2$.

 It is an amazing fact that the wrapping transition is controlled by two very disparate length scales: the scattering length $l$, and the system width $L_x$.   \footnote{The sensitivity to the system width was known to Altshuler and Sharvin, who included it in their calculation of the effect of a small perpendicular field on AAS oscillations. \cite{Sharvin84,aronov1987magnetic} } This is a consequence of the equipartition of energy between kinetic and potential energy.   In the Landau level problem the potential energy is  $ L_\phi^2 \langle r^2 \rangle /2 L_B^4$ and the kinetic energy is  $L_\phi^2 \langle k^2  \rangle / 2$, where $k$ is the wave-number and $r$ is the radius.  Setting the two equal, we obtain $ \langle k^2 \rangle = \langle r^2 \rangle / L_B^4$.  This evidences the linear relation between a state's momentum and radius, which is a special feature of the Landau level problem.     In other words,   states with large spatial extension also  have large momenta; their spatial  oscillations are faster.   This is why the wrapping transition is sensitive to the scattering length.  A change in scattering length translates directly to a corresponding change in the state's maximum spatial extent, which never exceeds  $L_B^2 / l$.   Comparing the state's extent to  $L_x/ 2$, we recover the equation for the wrapping transition, $L_B^2 = L_x l / 2$.
 
 % RMP v.59 p. 755 says that when there is a field component tilted off-axis in a a cylinder the dephasing length $1/\sqrt{L_\phi^2}$ gets updated to $\sqrt{1/L_\phi^2 + 2 R^2 / L_B^4}$.  In this connection it cites a review by Sharvin 1984: Sharvin, Y. V., 1984, Physica B 126, 288.  This seems roughly correct, mapping $a/\sqrt{L_\phi^2}$ $a R / L_B^2$. when $1 < R \sqrt{L_\phi^2}/L_B^2$.  In the review Sharvin says that this formula is correct when $L_B > R$ and that Altshuler pointed this physics out to him.

 The wrapping transition can be observed only if it occurs at a field strength which significantly exceeds the HLN shoulder at $L_B = L_\phi$; i.e. if $L_x l / 2 < L_\phi^2$.  Using $L_\phi^2 = D \tau_\phi = l v_F \tau_\phi / 2$, this condition simplifies to  $L_x  <  v_F \tau_\phi$, where $v_F$ is the Fermi velocity.  In other words,   the wrapping transition occurs only  when the width $L_x$ is smaller than $v_F \tau_\phi$, the length that the electron can move ballistically before dephasing.   This formula is very remarkable, for it reveals a ballistic  scale deep in the diffusive regime.

\subsection{Inside the Wrapping Transition}
Figure 4b focuses on the region inside the wrapping transition.  It shows the same data as Figure 4a, but in a log-log format which accentuates small  values of the magnetoconductivity and reveals a straight line when the signal follows a power law.  We can distinguish two regimes inside the wrapping transition.    First we discuss small fields, where we have plotted quadratic curves as straight lines without symbols.  The quadratic lines coincide very nicely with the signal,  which is always quadratic at small magnetic field.  The range of quadratic behavior is \textit{very roughly} $x< 4 L_\phi^2/ L_x^2$, which corresponds to values of the magnetic length $L_B^2$  exceeding $ L_x^2/2 $.  In this regime the magnetic field is not strong enough to create Cooperon bound states, and all Cooperon states wrap around the sample.  %small magnetoconductivities $\langle \Delta \sigma_{WAL}(x) \rangle < 0.2 $ the signal is always quadratic in the magnetic field.
   The observed quadratic signal is  similar to the HLN formula's quadratic regime where $\langle \Delta \sigma_{WAL}(x) \rangle = x^2/24$ when $L_B$ exceeds $L_\phi$, but here the signal is much  weaker and extends to much larger field strengths. 

Simple dimensional analysis shows that in the quadratic region where $\Delta \sigma_{WAL}(x) \rangle \propto B^2 \propto L_B^{-4}$, the magnetoconductivity must contain some length scale to the fourth power, to compensate for $B^2 \propto L_B^{-4}$.  In the HLN formula we have $L_\phi^4 / L_B^4$, while the in-plane magnetoconductivity is proportional to $L_y^2 L_\phi^2 / L_B^4$.  Our data in Figure 4b completely excludes  either type of scaling in the wrapped regime.  HLN scaling would produce an extra feature at $x = 1$, while scaling with $L_x^2 L_\phi^2 / L_B^4$ would produce  features  at values of $x = \left[2.5, 5, 10, 20\right]$.  Both types of scaling are absent.  Moreover, our analysis of the weak field data in Figure 4b at four values of the sample width $L_x$ indicates that the conductivity here is scaling with $ L_x^{3.5}$, and suggests that the conductivity scales as $\Delta \sigma_{WAL}(x) \rangle \propto L_x^{3.5} L_\phi^{0.5} L_B^{-4}$. We conclude that the conductivity's dependence on $L_\phi$ is strongly suppressed, and we will discuss this further in the next section.

In addition to the quadratic regime governed by wrapped states, Figure 4b shows also an intermediate  regime lying between the quadratic region at $L_B^2 \approx L_x^2/2$ and the wrapping transition at $L_B^2 = L_x l/2$.   Here the low energy Cooperon states do not make a full circuit around the TI, and remain restricted to one, two, or three faces.  Therefore the signal includes contributions not only from wrapped states, but also from the harmonic oscillator,  sidewall, and (in tilted fields) edge states which we have discussed earlier.  We analyze the harmonic oscillator and wrapped contributions in Appendix \ref{Analytical}.   Since this intermediate regime interpolates between  the quadratic signal seen at weak fields and the HLN logarithm at  strong fields,   in  this regime the magnetoconductivity will look roughly linear.  

        \subsection{Suppression of Sensitivity to $L_\phi$ and  to Temperature Inside the Wrapping Transition}
       The HLN formula is based on the assumption that the Cooperon is able to travel a distance $L_\phi$, where $L_\phi$ is associated with quantum decoherence  and typically scales with $L_\phi \propto T^{-1/2}$ in 2-D diffusive systems.   In this article we concern ourselves with  wires that are smaller than $L_\phi$, so that states are able to wrap around the wire's perimeter.   In this scenario the perimeter $L_r$ provides the fundamental upper limit on the Cooperon's travel distance, and its dependence on $L_\phi$ and on $T$ are suppressed.
       
Mathematically, the mechanism by which the $L_\phi$ dependence is suppressed in small samples originates with the level spacing of the eigenvalues of the Cooperon diffusion operator.  At $B = 0$ the energy levels    are $L_\phi^2 k_r^2 = n_r^2 (2 \pi L_\phi / L_r)^2$, with a level spacing of order $(2 \pi L_\phi / L_r)^2$.   This quantity should be added to the $1$ in the diffusion operator.    Therefore we adopt the following prescription for small samples: we replace $L_\phi^2$ with a new length scale $L_{\phi r}^2$, given by $L_{\phi r}^{-2} = L_\phi^{-2} + ( L_r/ 2\pi)^{-2}$.   This change of course does not affect the ultraviolet cutoff, which remains $E_l = L_\phi^2 / l^2$.  The net effect is that in small samples  $L_\phi$ should be replaced by $L_r/2 \pi$ which is independent of temperature, and in larger samples we use $L_{\phi r}$ which retains some weakened temperature dependence.

In appendix \ref{StraightPT} we perform an analytical analysis of the quadratic regime, which is governed by wrapped Cooperon states. We find that in this regime the conductivity is governed by two dimensionless constants.  The first constant $\gamma$ is determined by semiclassical considerations and is  equal to $\gamma = \langle R^2(r) \rangle \; L_{\phi r}^2 / L_B^4$, where $\langle R^2(r)\rangle $  is the second moment of $R(r)$ and is given in equation \ref{R2Formula}.   $\gamma$ scales with temperature as $\gamma \propto L_{\phi r}^2 $; it is independent of temperature in small samples and scales with $1/T$ in large samples.   The second dimensionless constant  is $\lambda^4/L_B^4$, where $\lambda \leq L_r$ is an explicitly quantum quantity that is independent of both $L_{\phi r}$ and temperature $T$.  $\lambda^4/L_B^4$ difficult to compute analytically.  It is never significantly greater than $\gamma$, and may be considerably smaller than $\gamma$.

   \begin{figure}[]
%\centering
\includegraphics[width=9cm,clip,angle=0]{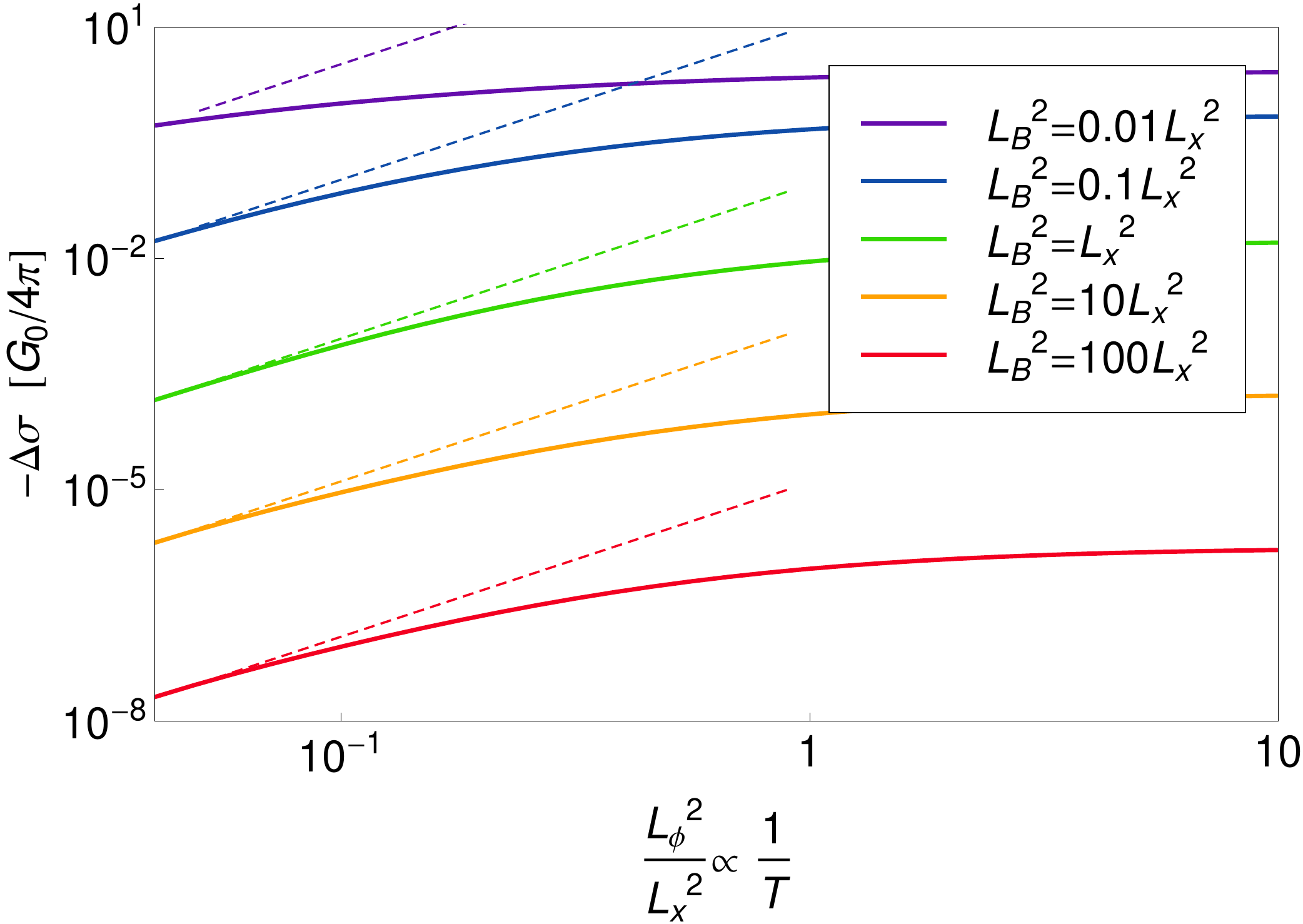}
\caption{ (Color online.)  Suppression of sensitivity to the dephasing length $L_\phi$ and the temperature $T$.     We plot the magnetoconductivity $\langle \sigma_{WAL} \rangle$ as a function of $L_\phi^2$, the square of the dephasing length, which is inversely proportional to the temperature $T$.  We keep the magnetic field strength fixed at  five values $L_B^2 = \left[ 0.01, 0.1, 1, 10, 100 \right] L_x^2$.   On this logarithmic graph the dashed lines show the scaling predicted by the HLN formula,  $ \sigma \propto L_\phi^4 \propto 1/ T^2$, and match the conductivity when  the coherence length is small $L_\phi \ll L_x$ and the magnetic field is also small.     When the coherence length is large (on the right side of the graph) the HLN formula fails and $\sigma$ is independent of both $L_\phi$ and $T$.   }
\label{FigAngle}
\end{figure}

If we omit $\lambda^4/L_B^4$ from our considerations, we obtain equation \ref{VCAConductivity} for the magnetoconductivity, which we repeat here:
\begin{eqnarray} \label{VCAConductivity1}
 \langle \Delta \sigma_{WAL}(B) \rangle
 & = & - \frac{G_0 }{4 \pi} \ln |1 + \gamma|, \; \gamma  = \frac{L_{\phi r}^2\langle R^2(r) \rangle }{ L_B^4}
\end{eqnarray}
This allows us to  reach some conclusions about the quadratic coefficient  $b$ which controls the conductivity at small fields via $\langle \Delta \sigma_{WAL}(B) \rangle = -(G_0/4\pi) \; b B^2$.   The HLN formula, which applies in large samples with large magnetic fields outside the wrapping transition, gives $b = 2 e^2 L_\phi^4 / 3 \hbar^2 \propto 1/ T^2$. In contrast, inside the wrapping transition we find  that
\begin{eqnarray}
b & \propto & \frac{4 e^2}{\hbar^2} \langle R^2(r) \rangle \; L_{\phi r}^2 
\end{eqnarray}
This $b$ scales with $L_r^2 L_\phi^2 \propto 1/T$ in samples that are much larger than $L_\phi$.  As $L_\phi$ is increased and surpasses the sample dimensions, the dependence on $L_\phi$ and temperature $T$ are suppressed, and $\sigma$  becomes completely independent of these quantities.  It scales instead with the sample dimension - $b \propto L_r^4$ - in samples that are small compared to $L_\phi$.  Figure 5 illustrates this effect at five values of the magnetic field strength corresponding to cyclotron radii $L_B$ which are smaller than, equal to, and greater than the sample size $L_x$.   The lower curves in Figure 5 show weak fields, i.e.  $L_B \gg L_x$, where the magnetoconductivity is quadratic in $B$.  Along the $x$ axis we vary the ratio of $L_\phi$ to the sample size.  On the left side of Figure 5 $L_\phi$ is small and our numerical results match well to the HLN $\sigma \propto L_\phi^4 \propto 1/ T^2$ curves.  As $L_\phi$ is increased our numerical curves progressively depart from the HLN prediction.  At $L_\phi \geq L_x$ the conductivity is roughly constant with respect to $L_\phi$; the temperature dependence is completely suppressed.

This effect will be easy to measure experimentally by reducing the temperature while keeping the magnetic field constant. It will manifest itself as a plateau in the magnetoconductivity at low temperatures where the dephasing length becomes comparable to the sample size.

  \subsection{ Changes in the sample height $L_y$.  }
      \begin{figure}[]
%\centering
\includegraphics[width=9cm,clip,angle=0]{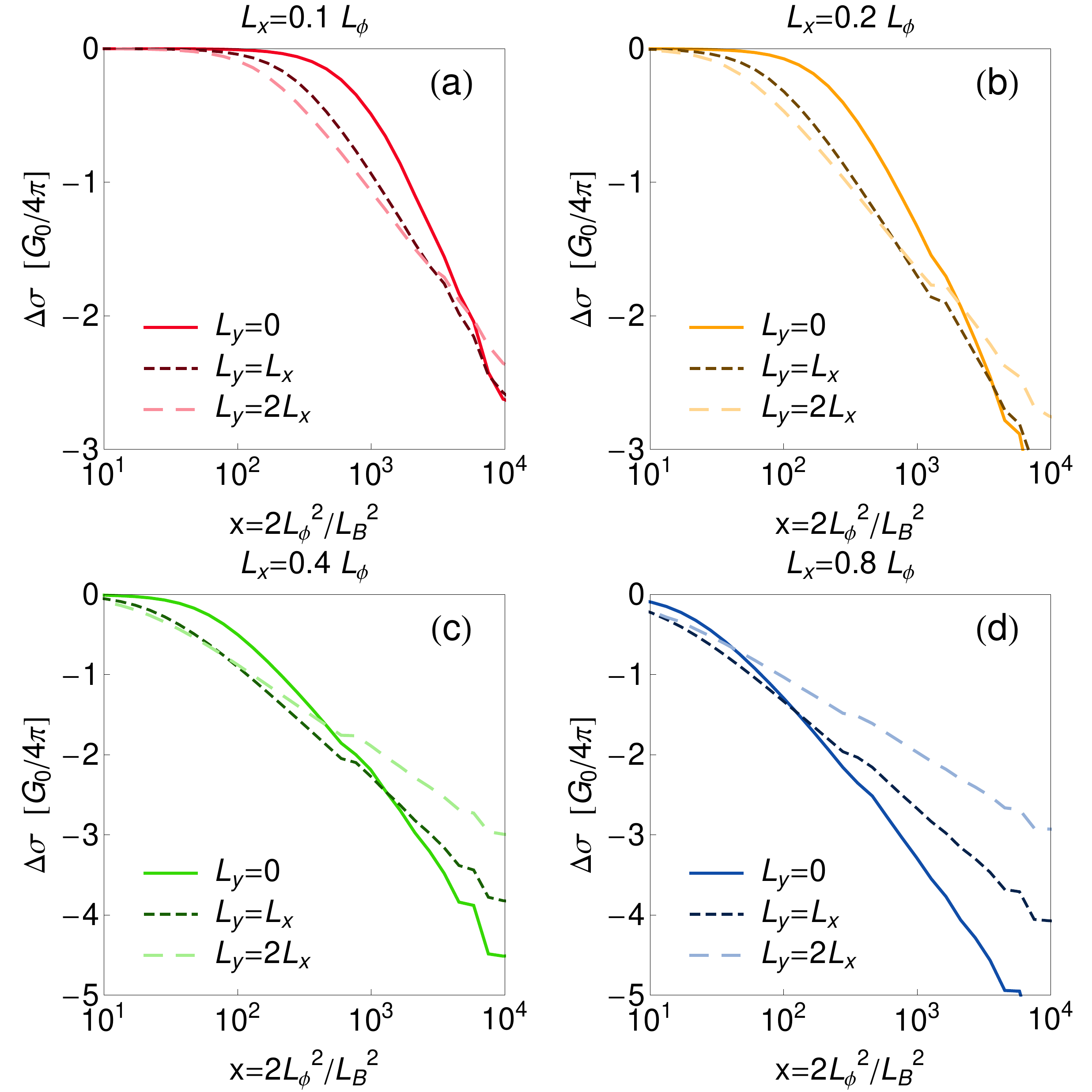}
\caption{ (Color online.)  The effect of sample height.  We plot the magnetoconductivity $\langle \sigma_{WAL}(x) \rangle$ as a function of $x \propto B$ at three aspect ratios $L_y = \left[0,  L_x, 2 L_x \right]$, signaled by unbroken lines, short dashes, and long dashes respectively.  We show samples with width $L_x = \left[0.8, 0.4, 0.2, 0.1 \right] L_\phi$.   Inside the wrapping transition $\langle \sigma_{WAL}(x) \rangle$ increases when $L_x$ increases, while the trend is reversed at large fields. }
\label{FigHeight}
\end{figure}
 Our previous results concerned infinitely thin TI samples.  This geometry  is a good approximation to thin films, which often have a width of microns and a depth of tens of nanometers.  Here we examine the magnetoconductivity of thicker samples.  The unbroken lines in Figure 6 show the zero-thickness results presented earlier, the lines with short dashes show square samples with height equal to width $L_y = L_x$, and the lines with long dashes show samples  where the height is twice the width $L_y = 2 L_x$. 
 
 Outside of the wrapping transition, at large fields, increasing the sample height causes the magnetoconductivity to decrease.  This is caused by the $ 2 L_x / L_r$ multiplier which we have discussed earlier.  While this multiplier is not accurate in small samples where $L_r \ll L_\phi$, it does become important when the sample height is increased, because the perimeter grows with sample height and eventually exceeds the dephasing length.

 Inside the wrapping transition, at small fields, the trend reverses; increasing the sample height causes  $\langle \sigma_{WAL}(B) \rangle $ to increase.  Consistently with this, the wrapping transition moves to smaller  field strengths.  This is an interesting signal of the TI's topology - if there were no transport along the sidewalls then the dependence on $L_y$ would be precisely null.   
 
 Quantitatively, the weak-field conductivity of an $L_x \times L_y = 0.1 L_\phi \times 0.1 L_\phi$ sample is a factor of $5.5$ larger than that of an $L_x \times L_y = 0.1 L_\phi \times 0$ sample.  When the height is doubled to $L_y = 0.2 L_\phi$, the conductivity is multiplied by an additional factor of $2.3$.  This $L_x \times L_y = 0.1 L_\phi \times 0.2 L_\phi$ conductivity  actually exceeds that of a wider but thinner $L_x \times L_y = 0.2 L_\phi \times 0$ sample.
 
  This effect and its sign  can be understood analytically in terms of the coefficient  $b$ which controls the conductivity at small fields, which we reported earlier:
\begin{eqnarray}
b & \propto & \frac{4 e^2}{\hbar^2} \langle R^2(r) \rangle \; L_{\phi r}^2
\end{eqnarray}
If the coherence length is much larger than the sample size $L_r / L_\phi \ll 1$, then $L_{\phi r} \approx L_r$, which doubles when the height is increased from zero to $L_y = L_x$.   The second moment $\langle R^2(r) \rangle$ is also doubled (see equation \ref{R2Formula}), so $b$'s net increase is a factor of $8$.  An additional height doubling to $L_y = 2 L_x$ causes $b$ to multiply by an additional factor of $2.625$.  Our numerical results are a bit less than these analytical predictions because we are not precisely in the $L_r / L_\phi \ll 1$ limit; instead we have $L_r / L_\phi = \left[ 0.2,\, 0. 4, \; 0.6 \right]$.

 %Our perturbative analysis in Appendix \ref{StraightPT} shows that the quadratic coefficient $b$  governing $\langle \sigma_{WAL}(B) \rangle = -(G_0 / 4 \pi) b B^2 $  at weak fields is determined by two dimensionless ratios.  The first is  $\gamma = L_\phi^2 \langle R^2(r) \rangle / L_B^4$, and the second is $\lambda^4 / L_B^4$.  $\lambda^4$ is difficult to evaluate, but Appendix  \ref{JWKB} shows that in a semiclassical approximation it contains one power of $\langle R^2(r) \rangle$.    A full treatment likely would show $\lambda^4$ is roughly proportional to the square of $\langle R^2(r) \rangle$.  In summary, $\langle R^2(r) \rangle$ sets the length scale of the magnetoconductivity; when it increases $\langle \sigma_{WAL}(B) \rangle $ also increases.

    \subsection{Tilted Magnetic Fields}
   \begin{figure}[]
%\centering
\includegraphics[width=8cm,clip,angle=0]{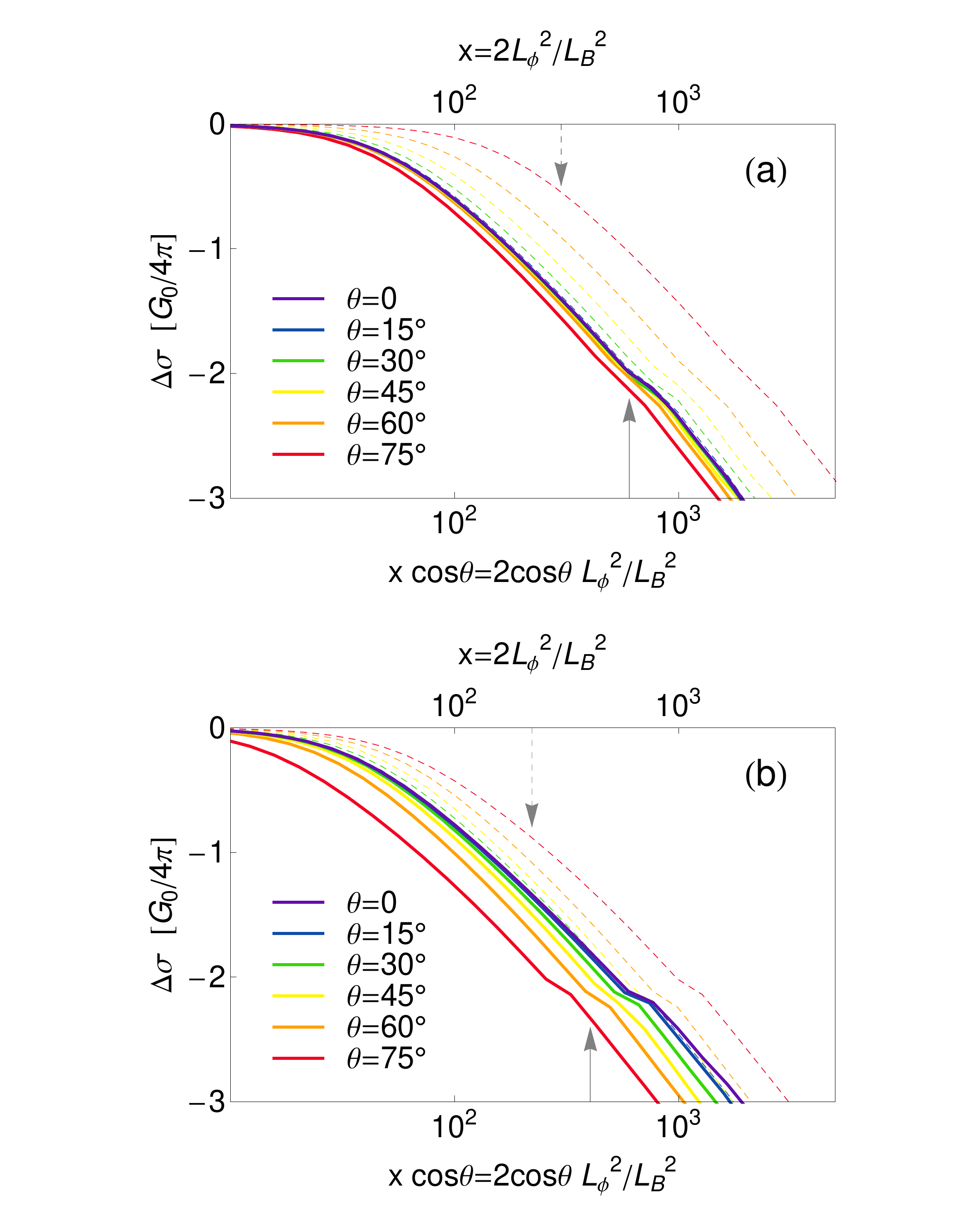}
\caption{ (Color online.)  Comparison between using the total magnetic field vs. its component perpendicular to the sample.  We plot the magnetoconductivity $\langle \sigma_{WAL}(x) \rangle$ at six field angles between perpendicular $\theta = 0$ and $\theta = 75$ degrees.  The unbroken lines are plotted with the perpendicular component $B \cos \theta$ as the $x$ axis, while the dashed lines are plotted with the total field $B$ as the $x$ axis.  In pane (a) we show a thin sample with $L_y = 0.1 L_x$, where use of the perpendicular field $B \cos \theta$ causes a nice collapse of the lines on top of each other, showing that $\langle \sigma_{WAL}(x) \rangle$  substantially depends only on $B \cos \theta$.  In pane (b) we show a thicker sample with $L_y = 0.4 L_x$, and find that the $B \cos \theta$ plot has no advantage over the $B$ plot. }
\label{FigAngle}
\end{figure}

Several experiments have reported tilted field measurements to confirm that in thin samples the observed magnetoconductivity is sensitive only to the perpendicular component of the magnetic field, $B \cos \theta$.   Using $B \cos \theta$ as the x-axis and $\Delta \sigma$ as the $y$ axis, these experiments showed agreement between measurements performed at different angles. \cite{chen2011tunable,He11, Wang12,Lee12,Cha12,Zhao13,Gehring13,Chen14} However many find that the agreement fails at large angles or large fields, and many also observe also an in-plane ($\theta = \pi /2$) signal which is inconsistent with dependence only on a perpendicular field.  \cite{He11,Cha12,Wang12,Zhao13,Chen14}  Here we present an analytical formula and numerical results on tilted fields. 

A simple formula which takes into account both in-plane and perpendicular fields can be obtained when the dephasing length $L_\phi$ is short compared to the sample size $L_r$.  In this case the magnetoconductivity is a sum of separate contributions from the TI's faces:   \begin{eqnarray}
\langle \sigma_{WAL}(B) \rangle  &=&  \frac{2L_x}{L_r} \sigma_{HLN}(x \cos \theta) + \frac{2L_y}{L_r} \sigma_{HLN}(x \sin \theta)  
\nonumber \\
\end{eqnarray}
Corrections are of order $O(L_\phi / L_r)$.  This same formula applies also to the regime of large magnetic fields, in which the cyclotron radius $\delta R$ is much smaller than the sample size.  This formula should be a distinct improvement over previous work which did not take into account the sidewall contribution $\frac{2L_y}{L_r} \sigma_{HLN}(x \sin \theta)  $.

Figure 7 shows our numerical results for tilted field measurements.  In pane 7a the width is $L_x = 0.4 L_\phi$, the height is one tenth of the width, and the scattering length is $l = L_\phi /100$.  We show results obtained at $\theta =0, 15, 30, 45, 60,$ and $75$ degrees.  To help the reader to make a comparison between plots using the perpendicular field $B \cos \theta$ and plots using the total field $B$, we plot both versions in the same graph.  The unbroken lines  are plotted using the perpendicular field $B \cos \theta$ as the $x$ axis, while  the dashed  lines show same data but are plotted using the total field $B$ as the $x$ axis.   

Figure 7a shows that in this thin $L_y = 0.1 L_x $ sample the data agrees well with a perpendicular field dependence: the lines with squares lie very near to each other, with the furthest departure at $75$ degrees.  In particular, the agreement between perpendicular-field plots (unbroken lines) is a large improvement  over the total-field plots (dashed lines), which proves that in this thin sample the perpendicular field, not the total field, controls the magnetoconductivity.  It is remarkable that the agreement is this good even though wrapping effects are important.  In fact the angular dependence shows  no qualitative difference between small field strengths inside the wrapping transition and large field strengths outside the wrapping transition.  

In contrast, in Figure 7b we show a thicker sample  where $L_y = 0.4 L_x$ and all other parameters remain the same.  Here using a perpendicular field gives no improvement over using the total field as the $x$ axis, presumably because the sidewall contribution is large.   Again we see little qualitative difference between results inside and outside the wrapping transition.  
     
   \begin{figure}[]
%\centering
    \includegraphics[width=8cm,clip,angle=0]{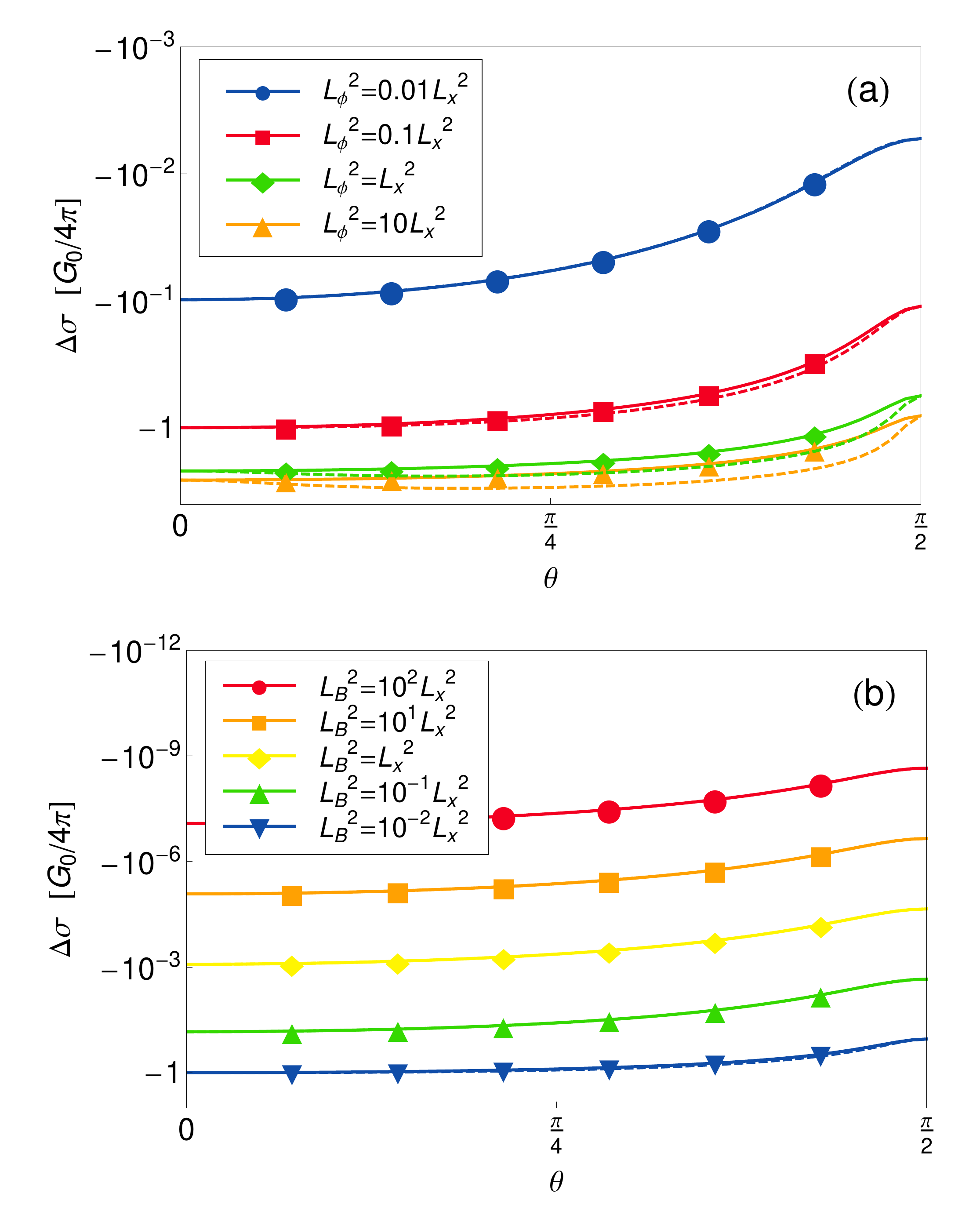}
\caption{ (Color online.) Dependence on the angle of the magnetic field in a thin sample with height $L_y = L_x / 10$.   $\theta = 0$ for perpendicular fields and $\theta = \pi/2$ for in-plane fields.  In panel a the magnetic field is kept fixed, and in panel b the dephasing length $L_\phi$ is kept fixed.  In both panels the upper curves lie in the strong-field regime.  For comparison purposes, the dashed lines show the sum of the perpendicular $\theta = 0$ and in-plane $\theta = \pi /2$ signals. In all cases the in-plane magnetoconductivity is smaller than the perpendicular value, with ratios ranging from $1/3$ to $1/40$.  This is in sharp contrast with a TI without sidewalls where the in-plane magnetoconductivity would be null, and may explain experimental measurements of in-plane signals. %todo: include the new compressed figure.
}
\label{FigTilt2}
\end{figure}

%  tilt2LB0p01DT0p1K300kyRes60(1)/tilt2LB0p01DT0p1K300kyRes60(50) =   37.3536
% tilt2LB0p1DT0p1K300kyRes60(1)/tilt2LB0p1DT0p1K300kyRes60(50) = 37.3528
%   tilt2LB1DT0p1K300kyRes60(1)/tilt2LB1DT0p1K300kyRes60(50) = 37.2938
% tilt2LB10DT0p1K300kyRes60(1)/tilt2LB10DT0p1K300kyRes60(50) =  31.1012
% tilt2LB100DT0p1K300kyRes60(1)/tilt2LB100DT0p1K300kyRes60(50) = 9.0119
% tilt2LB1000DT0p1K300kyRes60(1)/tilt2LB1000DT0p1K300kyRes60(50) =    5.9993

% tilt2LB100DT0p1K300kyRes60(1)/tilt2LB100DT0p1K300kyRes60(50) = 9.0119
% tilt2LB100DT1K300kyRes60(1)/tilt2LB100DT1K300kyRes60(50) = 3.8988
% tilt2LB100DT10K300kyRes60(1)/tilt2LB100DT10K300kyRes60(50) = 3.2018
% tilt2LB100DT100K300kyRes60(1)/tilt2LB100DT100K300kyRes60(50) =  3.1291

    Figure 8 shows our calculations of how the magnetoconductivity changes when the magnetic field angle is varied continuously between $\theta = 0$ (perpendicular to the TI sample) and $\theta = \pi / 2$ (parallel to the TI sample), a type of experiment which may prove useful for checking systematically the effects from the TI sidewalls.    In both panels the sample width is held fixed at $L_x = 1$, the sample is thin with height $L_y = L_x /10$, and the scattering length is $L_x/100$.  In panel 8a we keep the magnetic field fixed at $L_B = 0.1 L_x$ and choose four values of the coherence length $L_\phi^2 = \left[0.01, 0.1, 1, 10\right] L_x^2$.  In all cases we find that in-plane magnetoconductivity  $\sigma_{WAL}(B)$ is non-zero.  In fact the reduction from $\theta = 0$ to $\theta = \pi /2$ is only a factor of $9$ when $L_\phi = L_x / 10$, and a factor of $3$ when $L_\phi = \sqrt{10} L_x$.     Significantly, the factor of $9$ is close to the ratio of the sample's height to width $L_y / L_x = 10$.  In panel 8b  we keep $L_\phi = L_x / \sqrt{10}$ and move $L_B$ progressively from the strong field regime (at the bottom of the figure) to the weak field field regime (at the top), i.e $L_B = L_x \times \left [ \sqrt{0.01}, \sqrt{0.1}, 1, \sqrt{10}, \sqrt{100}\right]$.  Again the ratio of in-plane signal  to perpendicular signal is comparable to $L_y/L_x$, with values ranging from $40$ at large fields to $6$ at small fields. In the weak-field regime where $L_B \geq L_\phi = L_x / \sqrt{10}$, i.e. the upper four curves, we find that all four curves run in parallel.  This is because here the signal is proportional to $B^2$. The same parallel behavior is visible in the upper two curves of  panel 8a.  
    
    These numerical results may explain why sizable in-plane signals have been observed repeatedly in TI experiments.  Our results indicate that when $L_\phi / L_r$ is small the ratio of in-plane to perpendicular field signals is roughly $L_y/L_x$, and in the other cases the ratio is even larger.  When $L_\phi/L_x$ is large, i.e. in the wrapped regime, the ratio of in-plane to perpendicular signals can be quite large, since the relevant interference diagrams wrap around the entire TI sample.
    
    Figure 8 also shows dashed lines, which are the sum of the perpendicular conductivity $\sigma_{perp}(B \cos \theta)$ and the in-plane conductivity $\sigma_{in-plane}(B \sin \theta)$.  It is experimentally quite feasible to measure the perpendicular and in-plane conductivities and then compare $\sigma_{perp}(B \cos \theta) + \sigma_{in-plane}(B \sin \theta)$ to the tilted magnetoconductivity $\sigma(B, \theta)$.   When there is little coupling between the TI's four faces, the two curves should agree.  Figure 8a shows generally good agreement, with significant discrepancies only  when the coherence length $L_\phi$ is comparable to or larger than the sample size, precisely because in this case all four sides are coupled to each other.  The maximum discrepancy at  $L_\phi^2 = 0.01 L_x^2$ is $1.6\%$, at $L_\phi^2 = 0.1 L_x^2$ it grows to  $12\%$, and at $L_\phi^2 = 10 L_x^2$ it reaches $45\%$.  These discrepancies are small compared to $\sigma$'s overall variation with $\theta$.   Figure 8b also shows small discrepancies, less than $1\%$ at $L_B^2 \geq L_x^2 $, and growing to $13 \%$  at $L_B^2 = 0.01 L_x^2$.    We conclude that $\sigma_{perp}(B \cos \theta) + \sigma_{in-plane}(B \sin \theta)$ is a good predictor of the tilted magnetoconductivity, with almost perfect agreement when $L_\phi$ is small or $L_B$ is large.

\section{\label{Conclusions}Conclusions}
 In this article we have made a comprehensive exploration of topological effects on the Weak Antilocalization signal in TI samples with perpendicular, in-plane, and tilted magnetic fields.   These results are of interest because they bring to light a new experimentally accessible signal of topological transport.  They are also practically useful to experimentalists.  Our results will affect interpretation of the magnetoconductivity's magnitude, which is often used to estimate the coupling between a TI's top and bottom surfaces.  We also showed that TI sidewalls produce an in-plane magnetoconductivity which may account for experimental observations, 
 and  we gave both analytical and numerical predictions for tilted samples that explain experimental observations with in-plane fields.

\begin{acknowledgements}We acknowledge useful discussions with  Xi Dai, Yongqing Li, Chaojing Lin, Quansheng Wu, and Hu-Jong Lee.  K.B.A. and I.A.S. acknowledge the support of FP7 IRSES project QOCaN. This work was initiated at the Institute of Physics in Beijing,  with support from the National Science Foundation of China and the 973 program of China  under Contract No. 2011CBA00108.
\end{acknowledgements}

\begin{appendix}
\begin{widetext}

\section{Analytical Calculation of the Magnetoconductivity\label{Analytical}}
In this appendix we analytically calculate the magnetoconductivity.  In the first section we calculate the conductivity from wrapped Cooperon states, which is the full result at small fields. In our finite system the weak-field result is necessarily analytic and even in $B^2$, as we discuss in the next section.  Lastly we 
%develop a semiclassical approximation suitable for stronger fields  and use it to
 calculate the contribution from harmonic oscillator Cooperon states, which becomes important at stronger fields.

We begin by discussing the relevant length scales and expected results.
As discussed earlier, our numerical results show that any features scaling with $L_\phi$ are strongly suppressed or nonexistent in the small samples  of interest to us, where  wrapping is important.  This is because the wrapping effect requires that $L_\phi$ be larger than the sample perimeter.  As described earlier, we adopt the following prescription for small samples: we replace $L_\phi^2$ with a new length scale $L_{\phi r}^2$, given by $L_{\phi r}^{-2} = L_\phi^{-2} + ( L_r/ 2\pi)^{-2}$.   This has  two main results.  Firstly, in small $L_r \ll L_\phi$ samples the $L_\phi$ length scale is replaced by the sample size $L_r / 2\pi$. Secondly, in small samples the magnetoconductivity's dependence on temperature disappears.

\subsection{The Conductivity from Wrapped States - Equal to the Total Conductivity in Weak Fields\label{StraightPT}}

As shown in equation \ref{MCequation}, our calculation of the magnetoconductivity reduces to a study of the eigenstates and eigenvalues of the Cooperon diffusion operator $A$ which includes a potential representing the magnetic field:
\begin{eqnarray} \label{MCequationWeak}
\langle \sigma_{WAL}(B) \rangle
 & = &  \frac{G_0 }{4\pi}   \;2 L_{\phi r}^2 \; \int \frac{dk_z}{L_r} \; \sum_{i}^{E_i < E_l} E_i^{-1}, \; E_i | i \rangle =  A | i \rangle
 \nonumber \\
 A = & \;&   V(r, R_g) - L_{\phi r}^2 (\partial/\partial_r)^2  + 1, \; 
V(r, R_g) = (  L_{\phi r}^2 / L_B^4) (R(r) +L_B^2 k_z)^2
\end{eqnarray}
The function $R(r)$ figuring in the potential was specified earlier. The average value of $V(r,R_g)$ is $ \langle V \rangle = L_{\phi r}^2 k_z^2 +  \gamma, \; \gamma  = L_{\phi r}^2\langle R^2(r) \rangle / L_B^4$.  $\langle R^2(r) \rangle$ is the second moment of $R(r)$, scales quadratically with the perimeter $L_r$, and is given in equation \ref{R2Formula}.    

 In this section we concentrate on the contribution from wrapped Cooperon states, which can be analytically continued to $B=0$ where they are plane waves.  %Because of their analyticity, perturbation theory in powers of the potential is justified, and we develop this perturbation theory here.
In general, the conductivity includes contributions from wrapped, sidewall, edge, and harmonic oscillator states.  However at weak enough field strength the potential is not strong enough to generate any bound states, and all eigenstates and eigenvalues can be analytically continued to $B=0$; all Cooperon eigenstates are wrapped states.  (This is not true in an infinite system, where at any $B \neq 0$ the spectrum is broken into equally spaced Landau levels.)  In other words, at weak enough fields the potential $V(r,R_g)$ is not strong enough to create any bound states; therefore there is no contribution to the conductivity from sidewall, edge, or harmonic oscillator states.
 
The most obvious approximation for calculating the wrapped Cooperon  state contribution is the Virtual Crystal Approximation, \cite{VCA1, VCA2} which simply takes the average of the potential and obtains
\begin{eqnarray} \label{VCAEnergy}
E_i & = & 1 + L_{\phi r}^2 k_r^2 +  \langle V \rangle
\end{eqnarray}
$k_r  = n_r 2 \pi / L_r$ is the wave-number of the plane wave.   This approximation simply shifts the plane-wave eigenvalues by a dimensionless constant $\gamma = L_{\phi r}^2  \langle R^2(r) \rangle / L_B^4$, and the magnetic field's effect is controlled completely by this constant.  This approximation is closely linked to the magnetic phase relaxation approximation, which estimates the lowest excitation energy $\delta E$ of the Cooperon's ground state, and adds this $E_i$.  \cite{kettemann2002magnetolocalization,kettemann2007dimensional} The $\langle V \rangle$  term appearing in equation \ref{VCAEnergy} is the VCA approximation's estimate of the shift $\delta E$ of the Cooperon's  lowest excitation energy.

The remaining integral is easy because we have preserved rotational symmetry; we obtain 
\begin{eqnarray} \label{VCAConductivity}
\langle \sigma_{WAL}(B) \rangle
 & = &  \frac{G_0 }{4 \pi} (\ln | E_l | - \ln |1 + \gamma|)
 \nonumber \\
 \langle \Delta \sigma_{WAL}(B) \rangle
 & = & - \frac{G_0 }{4 \pi} \ln |1 + \gamma|, \; \gamma  = L_{\phi r}^2\langle R^2(r) \rangle / L_B^4
\end{eqnarray}

%todo: wrapped Cooperon states.  This notion might be confusing for some readers, we are at weak magnetic fields, where there are no Landau bands since the disorder smoothens them to a continous spectrum. Likewise the eigenstates of the Schršdinger equation are not HO eigenstates.  What we want to refer to is that the Cooperon equation Eq. 5 ff. has solutions E_i which would be for an infinite 2D system Landaulevels. For our system, the solutions are obtained by matching the 2D solutions with the side wall solutions of the Cooperon equation as You do numerically ( there should also be an analytical solution to this problem, but I havent tried yet). So, I guess it would be ok, if You susbtitute the naming ÒHO StateÓ by something else.

  \begin{figure}[]
%\centering
    \includegraphics[width=18cm,clip,angle=0]{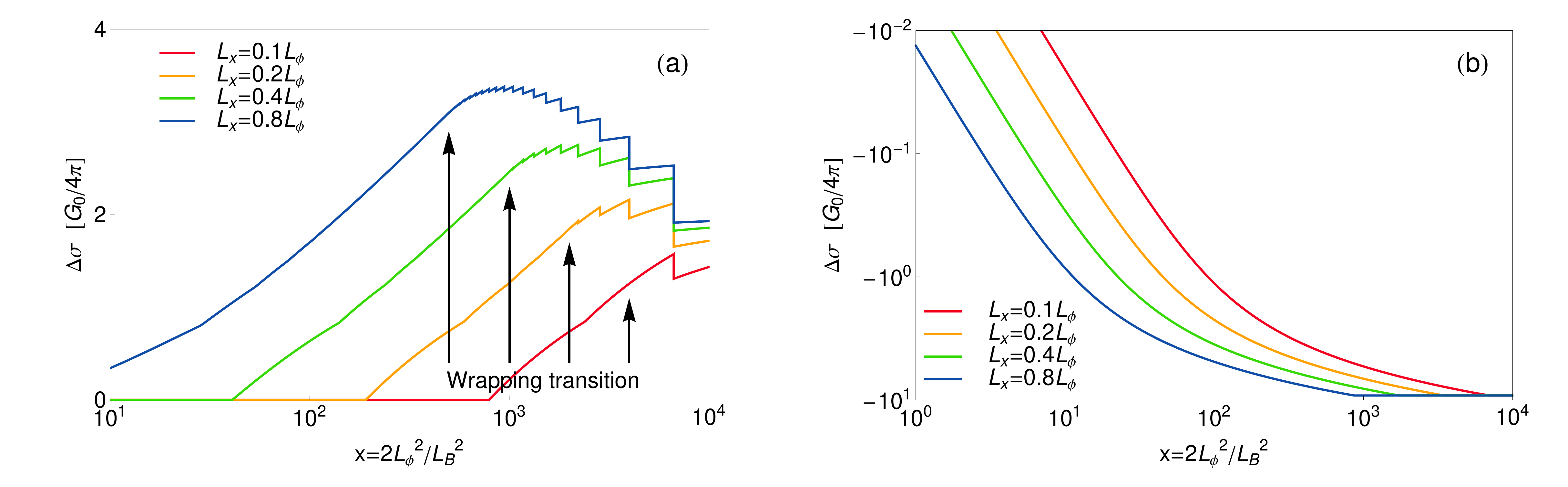}
\caption{ (Color online.) Analytical Theory of the Wrapped Magnetoconductivity. For easy comparison with Figure 4, here all parameters and scales are identical to those used in Figure 4. Pane a shows the contribution from harmonic oscillator bound states.  Inside the wrapping transition this contribution increases with $B$ because $B$ causes more bound states.  At the wrapping transition this trend reverses as the increasing bound state energies begin to cross the high-energy cutoff associated with the scattering length $l$.  The logarithmic decrease seen here at large fields is responsible for the identical feature in the HLN formula. Pane b shows the contribution from wrapped states.  At small fields we find the usual quadratic behavior, with a transition to logarithmic growth when  when $L_B $ is proportional to the sample size.  }
\label{FigTilt2}
\end{figure}

This formula is illustrated in Figure 9b, which uses identical parameters to those used in Figure 4. In the wrapped regime where the coherence length is much larger than the sample size $L_\phi \gg L_r$, the $\gamma$ parameter scales with $(L_r / L_B)^4$.  In this regime the magnetoconductivity is quadratic at small fields and transitions to logarithmic behavior at $L_B \propto L_r$.   Qualitatively this is the correct behavior.  However comparison with Figure 4 shows that at small fields the Virtual Crystal Approximation systematically overestimates the magnetoconductivity.  Moreover it does not capture the correct physics at field strengths which are intermediate between $L_B \propto L_r$ and the wrapping transition which occurs at  much bigger fields where $L_B \propto \sqrt{l L_r}$.  In this transition region the Virtual Crystal Approximation predicts logarithmic growth, while our numerical results show much faster growth, closer to a power law.  The reason for this discrepancy is that the Virtual Crystal Approximation neglects the conversion from wrapped states to bound states, i.e. to sidewall and harmonic oscillator states.

\subsubsection{Derivation of the Virtual Crystal Approximation from Perturbation Theory}
Here we do a correct perturbative expansion to second order in the magnetic field  $B \propto x \propto 1/L_B^2$.  At zeroth order we have $V(r, R_g) = L_{\phi r}^2 k_z^2$, and added to this is the perturbation $\delta V = 2 (L_{\phi r}^2 / L_B^2) k_z R(r) \, + \, L_{\phi r}^2 R^2(r) / L_B^4$.  Perturbation theory states that at second order in $V$ the energy is
\begin{eqnarray}
 E_i &=&1 + L_{\phi r}^2 k_z^2 + L_{\phi r}^2 k_r^2 + \langle k_r | \delta V | k_r \rangle + \sum_{j \neq n_r} \frac{|\langle k_r |\delta V | k_j \rangle|^2}{L_{\phi r}^2 (k_r^2 - k_j^2)}
\end{eqnarray}
If we omit the last term we obtain the Virtual Crystal Approximation, because $\langle n_r | \delta V | n_r \rangle = \gamma$. Keeping only terms of order $O(B^2)$ and simplifying produces the following result:
\begin{eqnarray}
E_i &=&1 +  L_{\phi r}^2 k_z^2 (1 + L_B^{-4} \lambda^4(k_r))  + L_{\phi r}^2 k_r^2 + \gamma, \; \lambda^4(k_r, L_x, L_y, l) =  4  \sum_{k_j \neq n_r} \frac{|\langle k_r | R(r)  | k_j \rangle|^2}{k_r^2 - k_j^2}
 \end{eqnarray}
This result  is exact up to second order in $B$, as long as the sample is finite.  Its most interesting feature  is the emergence of a new dimensionless parameter, $\lambda^4 / L_B^4$, which is completely independent of the dephasing length $L_{\phi r}$ and of temperature.  $\lambda^4$ cannot be larger than $O(L_r^4)$, and may be as small as $O(L_r^2 a^2)$.      We should note that the sign of $\lambda^4$ is not fixed a priori, and is guaranteed to be negative when $k_r = 0$; we used this notation only to point out the emergence of a new length scale.

This result leads us to expect that at small $B$ the magnetoconductivity is subject to competition between on one hand $L_{\phi r}^2 L_r^2 / L_B^4$ and on the other hand the  temperature-independent ratio  $\lambda^4 / L_B^4$.  In small samples the former scales with $L_r^4 / L_B^4$.  The latter  is bounded above by the same quantity, but may be as small as $L_r^2 a^2 / L_B^4$.

In the wrapped regime $\gamma \propto L_r^4$.  The $\lambda^4$ term cannot scale faster than this and may have a much slower scaling, which justifies its omission and the use of the Virtual Crystal Approximation. 

\subsection{Can the Magnetoconductivity Contain Terms that are Odd in $B$?}

It is of considerable interest to understand whether the magnetoconductivity may have terms that are linear in $B \propto x$ at small $B$.  In our equation \ref{MCequation} for the magnetoconductivity the integration $x \;dR_g$ is independent of $B$, and $x$ figures only in the potential $V(r,R_g)$.  Within the potential, the linear part is  is $-2 (x^2/4 L_{\phi r}^2) R_g \, R(r) =  x R(r) \; k_z$.  We will now show that if we do perturbation theory in this linear part, every term which is odd in $B \propto x$ will be exactly zero.

We obtain this result by noticing that the potential for our square TI sample shows a discrete symmetry: it is unchanged if it reverses sign and is shifted by $r \rightarrow r + L_x + L_y$.   Another way of saying this is that  the  $x R(r) \; k_z$ term mirrors itself - with opposite sign - under translation by $L_x + L_y$.  Since neither copy is preferred, odd powers of $x R(r)\; k_z$ must integrate to zero, and the magnetoconductivity cannot contain odd powers of $B$.  

More generally, in the original formulation with $ \vec{\nabla} - \imath 2 e \vec{A}/ \hbar$, if the geometry (without a magnetic field) retains symmetry under $ \vec{x} \rightarrow -\vec{x}$ then odd powers of $B \propto x$ are prohibited.

This argument does not exclude the possibility of non-analytic  terms which are linear in $B$,  associated with a failure of perturbation theory.  Such terms could appear if an infinitely small linear term generates new bound states that cannot be analytically continued to $B=0$, or if the density of states changes discontinuously at $B=0$.  In our problem with a simple magnetic field this can occur only if the system is infinite.  In that case the eigenvalues $E_n = |x| (n + 1/2)$ are those of the simple harmonic oscillator; they are linear in $B$, and are non-analytic when  $B$ changes sign.
 
\subsection{The Harmonic Oscillator Conductivity}
In the following we estimate the contribution from harmonic oscillator Cooperon states.  We assume that the magnetic field is perpendicular to the TI.     If the energy $E_n$ is small enough that the eigenfunction is trapped on one face of the TI, i.e. both turning points lie on that face, then the bound state eigenvalues coincide with those of the harmonic oscillator: $E^{HO}_n = 1 + (n + 1/2)  x |\frac{dR(r)}{dr} |$.  Therefore the conductivity is:
\begin{eqnarray}
\langle \sigma_{WAL}(B) \rangle & = &  2 \times G_0 \frac{L_{\phi r}^2  }{2\pi x}   \; \int_{-L_x/2L_B^2}^{L_x/2 L_B^2} \frac{dk_z}{L_r} \; \sum_{n = 0}^{1+(n+1/2)x< min(E_l, 1 + \xi^2 (1-|r|)^2/4)} (1/x  + (n + 1/2) )^{-1}
 \nonumber \\
 & = &   G_0 \frac{1 }{4\pi} \frac{2 L_x}{L_r} \; \int_0^{1 } {d r} \; \sum_{n = 0}^{1+(n+1/2)x< min(E_l, 1 + \xi^2 (1-r)^2/4)} (1/x  + (n + 1/2) )^{-1}
 \nonumber \\
 & = &   G_0 \frac{1 }{4\pi} \frac{2 L_x}{L_r}\; \sum_{n = 0}^{1+(n+1/2)x< min(E_l, 1 + \xi^2 /4)}  \; \int_0^{  1 - 2\sqrt{(n+1/2)x} /\xi } {d r}  \; (1/x  + (n + 1/2) )^{-1}
 \nonumber \\
 & = &   G_0 \frac{1 }{4\pi} \frac{2 L_x}{L_r}\; \sum_{n = 0}^{1+(n+1/2)x< min(E_l, 1 + \xi^2 /4)}    \; \frac{1 - 2\sqrt{(n+1/2)x} /\xi}{ 1/x  + n + 1/2 }
 \nonumber \\
 r &=& -2k_z L_B^2 /L_x, \; \xi =  L_{\phi r} L_x/ L_B^2 = x L_x / 2 L_{\phi r}
  \end{eqnarray}
  The factor of $2$ in the initial formula occurs because we are counting states from both the top and bottom faces of the TI.  The limits of integration $\pm L_x / 2 L_B^2$  are the values of $k_z$ where the quadratic sections of the potential $V(R_g, r)$ cease to have minima.  The factor of $ \xi^2 (1-|r|)^2/4$ in the sum's upper limit is the lower of $V(R_g, r)$'s two values at the sidewalls where $\pm L_x/2$.
 
 This formula is plotted in Figure 9a, with the same parameters used in our numerical calculations shown in Figure 4.  At large fields, i.e. outside the wrapping transition, all states are harmonic oscillator and sidewall bound states, and in thin samples the harmonic oscillator states are dominant. Therefore our harmonic oscillator formula is nearly exact at large fields, and in particular reproduces the correct logarithmic decrease.   At smaller fields below the wrapping transition at $L_B \propto \sqrt{L_r l}$ the potential becomes too weak to trap any bound states, and the harmonic oscillator contribution disappears, as shown in Figure 9a.   Our formula also shows the position of the wrapping transition.

The physics at smaller fields inside the wrapping transition involves a nonperturbative transformation of the states and their spectrum from harmonic oscillator states, to sidewall bound states, to finally unbound wrapped states which can be analytically continued to plane waves.  Our formula omits all details of this process, so our formula is inaccurate inside the wrapping transition.  Neither the Virtual Crystal Approximation nor the harmonic oscillator contribution, nor their combination, is able to give a qualitatively adequate picture of intermediate field strengths between $L_B \propto L_r$ and $L_B \propto \sqrt{L_\phi L_r}$.

 \subsection{Summary of our Analytical Results}
In this appendix we have calculated the contributions from wrapped  and harmonic oscillator Cooperon states to the  quantum conductivity $\langle \sigma_{WAL} (B) \rangle$.  The harmonic oscillator formula shows the wrapping transition and gives the correct strong-field logarithmic behavior.
%The wrapped contribution is even in $B$, but the sidewall and harmonic oscillator contributions contain terms which are odd in $B$ and are directly caused by the sample's finite size. 
At small fields the wrapped contribution  is equal to the total signal.  We developed the Virtual Crystal Approximation which gives a qualitatively correct picture of the weak-field quadratic region, and of the transition at stronger fields $L_B^2 < L_x^2/2$.   It also informed us of the signal's dependence on $ \gamma  = L_{\phi r}^2\langle R^2(r) \rangle / L_B^4$, which  this article's main body showed   is important for understanding the effects of changing the sample height, and of changing the temperature. 
% We also made an attempt at going beyond the Virtual Crystal Approximation, which produced only qualitative information about a second dimensionless parameter $\lambda^4/L_B^4$.

%-AB ring in ballistic regime, with no magnetic field and fully symmetric field and no spin orbit coupling, zero magnetic field gives perfect constructive interference and  when you turn on a magnetic field then conductance goes down.  If you have strong spin-orbit coupling then the conductance goes up because there is an effective magnetic field which is perpendicular to the axis of the ring.

\end{widetext}
\end{appendix}
\bibliography{Vincent}

\begin{thebibliography}{51}
\expandafter\ifx\csname natexlab\endcsname\relax\def\natexlab#1{#1}\fi
\expandafter\ifx\csname bibnamefont\endcsname\relax
  \def\bibnamefont#1{#1}\fi
\expandafter\ifx\csname bibfnamefont\endcsname\relax
  \def\bibfnamefont#1{#1}\fi
\expandafter\ifx\csname citenamefont\endcsname\relax
  \def\citenamefont#1{#1}\fi
\expandafter\ifx\csname url\endcsname\relax
  \def\url#1{\texttt{#1}}\fi
\expandafter\ifx\csname urlprefix\endcsname\relax\def\urlprefix{URL }\fi
\providecommand{\bibinfo}[2]{#2}
\providecommand{\eprint}[2][]{\url{#2}}

\bibitem[{\citenamefont{Kane and Mele}(2005)}]{Kane05}
\bibinfo{author}{\bibfnamefont{C.~L.} \bibnamefont{Kane}} \bibnamefont{and}
  \bibinfo{author}{\bibfnamefont{E.~J.} \bibnamefont{Mele}},
  \bibinfo{journal}{Physical Review Letters} \textbf{\bibinfo{volume}{95}},
  \bibinfo{pages}{226801} (\bibinfo{year}{2005}).

\bibitem[{\citenamefont{Min et~al.}(2006)\citenamefont{Min, Hill, Sinitsyn,
  Sahu, Kleinman, and MacDonald}}]{Min06}
\bibinfo{author}{\bibfnamefont{H.}~\bibnamefont{Min}},
  \bibinfo{author}{\bibfnamefont{J.~E.} \bibnamefont{Hill}},
  \bibinfo{author}{\bibfnamefont{N.~A.} \bibnamefont{Sinitsyn}},
  \bibinfo{author}{\bibfnamefont{B.~R.} \bibnamefont{Sahu}},
  \bibinfo{author}{\bibfnamefont{L.}~\bibnamefont{Kleinman}}, \bibnamefont{and}
  \bibinfo{author}{\bibfnamefont{A.~H.} \bibnamefont{MacDonald}},
  \bibinfo{journal}{Physical Review B} \textbf{\bibinfo{volume}{74}},
  \bibinfo{pages}{165310} (\bibinfo{year}{2006}).

\bibitem[{\citenamefont{Zhang et~al.}(2009)\citenamefont{Zhang, Liu, Qi, Dai,
  Fang, and Zhang}}]{Zhang09}
\bibinfo{author}{\bibfnamefont{H.}~\bibnamefont{Zhang}},
  \bibinfo{author}{\bibfnamefont{C.-X.} \bibnamefont{Liu}},
  \bibinfo{author}{\bibfnamefont{X.-L.} \bibnamefont{Qi}},
  \bibinfo{author}{\bibfnamefont{X.}~\bibnamefont{Dai}},
  \bibinfo{author}{\bibfnamefont{Z.}~\bibnamefont{Fang}}, \bibnamefont{and}
  \bibinfo{author}{\bibfnamefont{S.-C.} \bibnamefont{Zhang}},
  \bibinfo{journal}{Nature Physics} \textbf{\bibinfo{volume}{5}},
  \bibinfo{pages}{438} (\bibinfo{year}{2009}).

\bibitem[{\citenamefont{Hasan and Kane}(2010)}]{Hasan10}
\bibinfo{author}{\bibfnamefont{M.~Z.} \bibnamefont{Hasan}} \bibnamefont{and}
  \bibinfo{author}{\bibfnamefont{C.~L.} \bibnamefont{Kane}},
  \bibinfo{journal}{Reviews of Modern Physics} \textbf{\bibinfo{volume}{82}},
  \bibinfo{pages}{3045} (\bibinfo{year}{2010}).

\bibitem[{\citenamefont{Culcer}(2012)}]{Culcer12}
\bibinfo{author}{\bibfnamefont{D.}~\bibnamefont{Culcer}},
  \bibinfo{journal}{Physica E} \textbf{\bibinfo{volume}{44}},
  \bibinfo{pages}{860} (\bibinfo{year}{2012}).

\bibitem[{\citenamefont{Bardarson and Moore}(2013)}]{Bardarson13}
\bibinfo{author}{\bibfnamefont{J.~H.} \bibnamefont{Bardarson}}
  \bibnamefont{and} \bibinfo{author}{\bibfnamefont{J.~E.} \bibnamefont{Moore}},
  \bibinfo{journal}{Reports on Progress in Phyiscs}
  \textbf{\bibinfo{volume}{76}}, \bibinfo{pages}{056501}
  (\bibinfo{year}{2013}).

\bibitem[{\citenamefont{Ostrovsky et~al.}(2007)\citenamefont{Ostrovsky, Gornyi,
  and Mirlin}}]{Ostrovsky07}
\bibinfo{author}{\bibfnamefont{P.~M.} \bibnamefont{Ostrovsky}},
  \bibinfo{author}{\bibfnamefont{I.~V.} \bibnamefont{Gornyi}},
  \bibnamefont{and} \bibinfo{author}{\bibfnamefont{A.~D.}
  \bibnamefont{Mirlin}}, \bibinfo{journal}{Physical Review Letters}
  \textbf{\bibinfo{volume}{98}}, \bibinfo{pages}{256801}
  (\bibinfo{year}{2007}).

\bibitem[{\citenamefont{Nomura et~al.}(2007)\citenamefont{Nomura, Koshino, and
  Ryu}}]{Nomura07}
\bibinfo{author}{\bibfnamefont{K.}~\bibnamefont{Nomura}},
  \bibinfo{author}{\bibfnamefont{M.}~\bibnamefont{Koshino}}, \bibnamefont{and}
  \bibinfo{author}{\bibfnamefont{S.}~\bibnamefont{Ryu}},
  \bibinfo{journal}{Physical Review Letters} \textbf{\bibinfo{volume}{99}},
  \bibinfo{pages}{146806} (\bibinfo{year}{2007}).

\bibitem[{\citenamefont{Mucciolo and Lewenkopf}(2010)}]{Mucciolo10}
\bibinfo{author}{\bibfnamefont{E.~R.} \bibnamefont{Mucciolo}} \bibnamefont{and}
  \bibinfo{author}{\bibfnamefont{C.~H.} \bibnamefont{Lewenkopf}},
  \bibinfo{journal}{J. Phys.: Condens. Matter} \textbf{\bibinfo{volume}{22}},
  \bibinfo{pages}{273201} (\bibinfo{year}{2010}).

\bibitem[{\citenamefont{Wu and Sacksteder}(2013)}]{wu2013effects}
\bibinfo{author}{\bibfnamefont{Q.}~\bibnamefont{Wu}} \bibnamefont{and}
  \bibinfo{author}{\bibfnamefont{V.~E.} \bibnamefont{Sacksteder}},
  \bibinfo{journal}{arXiv preprint arXiv:1309.6738}  (\bibinfo{year}{2013}).

\bibitem[{\citenamefont{Chen et~al.}(2010)\citenamefont{Chen, Qin, Yang, Liu,
  Guan, Qu, Zhang, Shi, Xie, Yang et~al.}}]{Chen10}
\bibinfo{author}{\bibfnamefont{J.}~\bibnamefont{Chen}},
  \bibinfo{author}{\bibfnamefont{H.~J.} \bibnamefont{Qin}},
  \bibinfo{author}{\bibfnamefont{F.}~\bibnamefont{Yang}},
  \bibinfo{author}{\bibfnamefont{J.}~\bibnamefont{Liu}},
  \bibinfo{author}{\bibfnamefont{T.}~\bibnamefont{Guan}},
  \bibinfo{author}{\bibfnamefont{F.~M.} \bibnamefont{Qu}},
  \bibinfo{author}{\bibfnamefont{G.~H.} \bibnamefont{Zhang}},
  \bibinfo{author}{\bibfnamefont{J.~R.} \bibnamefont{Shi}},
  \bibinfo{author}{\bibfnamefont{X.~C.} \bibnamefont{Xie}},
  \bibinfo{author}{\bibfnamefont{C.~L.} \bibnamefont{Yang}},
  \bibnamefont{et~al.}, \bibinfo{journal}{Physical Review Letters}
  \textbf{\bibinfo{volume}{105}}, \bibinfo{pages}{176602}
  (\bibinfo{year}{2010}).

\bibitem[{\citenamefont{Li et~al.}(2012)\citenamefont{Li, Wu, Shi, and
  Xie}}]{Li12}
\bibinfo{author}{\bibfnamefont{Y.~Q.} \bibnamefont{Li}},
  \bibinfo{author}{\bibfnamefont{K.~H.} \bibnamefont{Wu}},
  \bibinfo{author}{\bibfnamefont{J.~R.} \bibnamefont{Shi}}, \bibnamefont{and}
  \bibinfo{author}{\bibfnamefont{X.~C.} \bibnamefont{Xie}},
  \bibinfo{journal}{Front. Phys.} \textbf{\bibinfo{volume}{7}},
  \bibinfo{pages}{165} (\bibinfo{year}{2012}).

\bibitem[{\citenamefont{Hikami et~al.}(1980)\citenamefont{Hikami, Larkin, and
  Nagaoka}}]{Hikami80}
\bibinfo{author}{\bibfnamefont{S.}~\bibnamefont{Hikami}},
  \bibinfo{author}{\bibfnamefont{A.~I.} \bibnamefont{Larkin}},
  \bibnamefont{and} \bibinfo{author}{\bibfnamefont{Y.}~\bibnamefont{Nagaoka}},
  \bibinfo{journal}{Prog. Theor. Phys. Progress Letters}
  \textbf{\bibinfo{volume}{63}}, \bibinfo{pages}{707} (\bibinfo{year}{1980}).

\bibitem[{\citenamefont{Chen et~al.}(2011)\citenamefont{Chen, He, Wu, Ji, Lu,
  Shi, Smet, and Li}}]{chen2011tunable}
\bibinfo{author}{\bibfnamefont{J.}~\bibnamefont{Chen}},
  \bibinfo{author}{\bibfnamefont{X.~Y.} \bibnamefont{He}},
  \bibinfo{author}{\bibfnamefont{K.~H.} \bibnamefont{Wu}},
  \bibinfo{author}{\bibfnamefont{Z.~Q.} \bibnamefont{Ji}},
  \bibinfo{author}{\bibfnamefont{L.}~\bibnamefont{Lu}},
  \bibinfo{author}{\bibfnamefont{J.~R.} \bibnamefont{Shi}},
  \bibinfo{author}{\bibfnamefont{J.~H.} \bibnamefont{Smet}}, \bibnamefont{and}
  \bibinfo{author}{\bibfnamefont{Y.~Q.} \bibnamefont{Li}},
  \bibinfo{journal}{Physical Review B} \textbf{\bibinfo{volume}{83}},
  \bibinfo{pages}{241304} (\bibinfo{year}{2011}).

\bibitem[{\citenamefont{Steinberg et~al.}(2011)\citenamefont{Steinberg,
  Lalo{\"e}, Fatemi, Moodera, and Jarillo-Herrero}}]{steinberg2011electrically}
\bibinfo{author}{\bibfnamefont{H.}~\bibnamefont{Steinberg}},
  \bibinfo{author}{\bibfnamefont{J.-B.} \bibnamefont{Lalo{\"e}}},
  \bibinfo{author}{\bibfnamefont{V.}~\bibnamefont{Fatemi}},
  \bibinfo{author}{\bibfnamefont{J.~S.} \bibnamefont{Moodera}},
  \bibnamefont{and}
  \bibinfo{author}{\bibfnamefont{P.}~\bibnamefont{Jarillo-Herrero}},
  \bibinfo{journal}{Physical Review B} \textbf{\bibinfo{volume}{84}},
  \bibinfo{pages}{233101} (\bibinfo{year}{2011}).

\bibitem[{\citenamefont{Garate and Glazman}(2012)}]{garate2012weak}
\bibinfo{author}{\bibfnamefont{I.}~\bibnamefont{Garate}} \bibnamefont{and}
  \bibinfo{author}{\bibfnamefont{L.}~\bibnamefont{Glazman}},
  \bibinfo{journal}{Physical Review B} \textbf{\bibinfo{volume}{86}},
  \bibinfo{pages}{035422} (\bibinfo{year}{2012}).

\bibitem[{\citenamefont{Al'tshuler et~al.}(1981)\citenamefont{Al'tshuler,
  Aronov, and Spivak}}]{aas81}
\bibinfo{author}{\bibfnamefont{B.}~\bibnamefont{Al'tshuler}},
  \bibinfo{author}{\bibfnamefont{A.}~\bibnamefont{Aronov}}, \bibnamefont{and}
  \bibinfo{author}{\bibfnamefont{B.}~\bibnamefont{Spivak}},
  \bibinfo{journal}{JETP Lett} \textbf{\bibinfo{volume}{33}},
  \bibinfo{pages}{94} (\bibinfo{year}{1981}).

\bibitem[{\citenamefont{Aronov and Sharvin}(1987)}]{aronov1987magnetic}
\bibinfo{author}{\bibfnamefont{A.}~\bibnamefont{Aronov}} \bibnamefont{and}
  \bibinfo{author}{\bibfnamefont{Y.~V.} \bibnamefont{Sharvin}},
  \bibinfo{journal}{Reviews of modern physics} \textbf{\bibinfo{volume}{59}},
  \bibinfo{pages}{755} (\bibinfo{year}{1987}).

\bibitem[{\citenamefont{Peng et~al.}(2010)\citenamefont{Peng, Lai, Kong,
  Meister, Chen, Qi, Zhang, Shen, and Cui}}]{peng2010aharonov}
\bibinfo{author}{\bibfnamefont{H.}~\bibnamefont{Peng}},
  \bibinfo{author}{\bibfnamefont{K.}~\bibnamefont{Lai}},
  \bibinfo{author}{\bibfnamefont{D.}~\bibnamefont{Kong}},
  \bibinfo{author}{\bibfnamefont{S.}~\bibnamefont{Meister}},
  \bibinfo{author}{\bibfnamefont{Y.}~\bibnamefont{Chen}},
  \bibinfo{author}{\bibfnamefont{X.-L.} \bibnamefont{Qi}},
  \bibinfo{author}{\bibfnamefont{S.-C.} \bibnamefont{Zhang}},
  \bibinfo{author}{\bibfnamefont{Z.-X.} \bibnamefont{Shen}}, \bibnamefont{and}
  \bibinfo{author}{\bibfnamefont{Y.}~\bibnamefont{Cui}},
  \bibinfo{journal}{Nature materials} \textbf{\bibinfo{volume}{9}},
  \bibinfo{pages}{225} (\bibinfo{year}{2010}).

\bibitem[{\citenamefont{Xiu et~al.}(2011)\citenamefont{Xiu, He, Wang, Cheng,
  Chang, Lang, Huang, Kou, Zhou, Jiang et~al.}}]{xiu2011manipulating}
\bibinfo{author}{\bibfnamefont{F.}~\bibnamefont{Xiu}},
  \bibinfo{author}{\bibfnamefont{L.}~\bibnamefont{He}},
  \bibinfo{author}{\bibfnamefont{Y.}~\bibnamefont{Wang}},
  \bibinfo{author}{\bibfnamefont{L.}~\bibnamefont{Cheng}},
  \bibinfo{author}{\bibfnamefont{L.-T.} \bibnamefont{Chang}},
  \bibinfo{author}{\bibfnamefont{M.}~\bibnamefont{Lang}},
  \bibinfo{author}{\bibfnamefont{G.}~\bibnamefont{Huang}},
  \bibinfo{author}{\bibfnamefont{X.}~\bibnamefont{Kou}},
  \bibinfo{author}{\bibfnamefont{Y.}~\bibnamefont{Zhou}},
  \bibinfo{author}{\bibfnamefont{X.}~\bibnamefont{Jiang}},
  \bibnamefont{et~al.}, \bibinfo{journal}{Nature nanotechnology}
  \textbf{\bibinfo{volume}{6}}, \bibinfo{pages}{216} (\bibinfo{year}{2011}).

\bibitem[{\citenamefont{Sulaev et~al.}(2013)\citenamefont{Sulaev, Ren, Xia,
  Lin, Yu, Qiu, Zhang, Han, Li, Zhu et~al.}}]{sulaev2013experimental}
\bibinfo{author}{\bibfnamefont{A.}~\bibnamefont{Sulaev}},
  \bibinfo{author}{\bibfnamefont{P.}~\bibnamefont{Ren}},
  \bibinfo{author}{\bibfnamefont{B.}~\bibnamefont{Xia}},
  \bibinfo{author}{\bibfnamefont{Q.~H.} \bibnamefont{Lin}},
  \bibinfo{author}{\bibfnamefont{T.}~\bibnamefont{Yu}},
  \bibinfo{author}{\bibfnamefont{C.}~\bibnamefont{Qiu}},
  \bibinfo{author}{\bibfnamefont{S.-Y.} \bibnamefont{Zhang}},
  \bibinfo{author}{\bibfnamefont{M.-Y.} \bibnamefont{Han}},
  \bibinfo{author}{\bibfnamefont{Z.~P.} \bibnamefont{Li}},
  \bibinfo{author}{\bibfnamefont{W.~G.} \bibnamefont{Zhu}},
  \bibnamefont{et~al.} (\bibinfo{year}{2013}).

\bibitem[{\citenamefont{Dufouleur et~al.}(2013)\citenamefont{Dufouleur, Veyrat,
  Teichgr{\"a}ber, Neuhaus, Nowka, Hampel, Cayssol, Schumann, Eichler, Schmidt
  et~al.}}]{dufouleur2013quasiballistic}
\bibinfo{author}{\bibfnamefont{J.}~\bibnamefont{Dufouleur}},
  \bibinfo{author}{\bibfnamefont{L.}~\bibnamefont{Veyrat}},
  \bibinfo{author}{\bibfnamefont{A.}~\bibnamefont{Teichgr{\"a}ber}},
  \bibinfo{author}{\bibfnamefont{S.}~\bibnamefont{Neuhaus}},
  \bibinfo{author}{\bibfnamefont{C.}~\bibnamefont{Nowka}},
  \bibinfo{author}{\bibfnamefont{S.}~\bibnamefont{Hampel}},
  \bibinfo{author}{\bibfnamefont{J.}~\bibnamefont{Cayssol}},
  \bibinfo{author}{\bibfnamefont{J.}~\bibnamefont{Schumann}},
  \bibinfo{author}{\bibfnamefont{B.}~\bibnamefont{Eichler}},
  \bibinfo{author}{\bibfnamefont{O.}~\bibnamefont{Schmidt}},
  \bibnamefont{et~al.}, \bibinfo{journal}{Physical review letters}
  \textbf{\bibinfo{volume}{110}}, \bibinfo{pages}{186806}
  (\bibinfo{year}{2013}).

\bibitem[{\citenamefont{Altshuler and Aronov}(1981)}]{Altshuler81}
\bibinfo{author}{\bibfnamefont{B.~L.} \bibnamefont{Altshuler}}
  \bibnamefont{and} \bibinfo{author}{\bibfnamefont{A.~G.}
  \bibnamefont{Aronov}}, \bibinfo{journal}{JETP Letters}
  \textbf{\bibinfo{volume}{33}}, \bibinfo{pages}{499} (\bibinfo{year}{1981}).

\bibitem[{\citenamefont{Dugaev and
  Khrnel'nitskii}(1984)}]{dugaev1984magnetoresistance}
\bibinfo{author}{\bibfnamefont{V.}~\bibnamefont{Dugaev}} \bibnamefont{and}
  \bibinfo{author}{\bibfnamefont{D.}~\bibnamefont{Khrnel'nitskii}},
  \bibinfo{journal}{Zh. Eksp. Teor. Fiz} \textbf{\bibinfo{volume}{86}},
  \bibinfo{pages}{1784} (\bibinfo{year}{1984}).

\bibitem[{\citenamefont{Beenakker and
  Van~Houten}(1988)}]{beenakker1988boundary}
\bibinfo{author}{\bibfnamefont{C.~W.~J.} \bibnamefont{Beenakker}}
  \bibnamefont{and}
  \bibinfo{author}{\bibfnamefont{H.}~\bibnamefont{Van~Houten}},
  \bibinfo{journal}{Physical Review B} \textbf{\bibinfo{volume}{38}},
  \bibinfo{pages}{3232} (\bibinfo{year}{1988}).

\bibitem[{\citenamefont{Tkachov and Hankiewicz}(2013)}]{tkachov2013spin}
\bibinfo{author}{\bibfnamefont{G.}~\bibnamefont{Tkachov}} \bibnamefont{and}
  \bibinfo{author}{\bibfnamefont{E.}~\bibnamefont{Hankiewicz}},
  \bibinfo{journal}{physica status solidi (b)} \textbf{\bibinfo{volume}{250}},
  \bibinfo{pages}{215} (\bibinfo{year}{2013}).

\bibitem[{\citenamefont{Raichev and Vasilopoulos}(2000)}]{raichev2000weak}
\bibinfo{author}{\bibfnamefont{O.}~\bibnamefont{Raichev}} \bibnamefont{and}
  \bibinfo{author}{\bibfnamefont{P.}~\bibnamefont{Vasilopoulos}},
  \bibinfo{journal}{Journal of Physics: Condensed Matter}
  \textbf{\bibinfo{volume}{12}}, \bibinfo{pages}{589} (\bibinfo{year}{2000}).

\bibitem[{\citenamefont{He et~al.}(2011)\citenamefont{He, Wang, Zhang, Sou,
  Wong, Wang, Lu, Shen, and Zhang}}]{He11}
\bibinfo{author}{\bibfnamefont{H.-T.} \bibnamefont{He}},
  \bibinfo{author}{\bibfnamefont{G.}~\bibnamefont{Wang}},
  \bibinfo{author}{\bibfnamefont{T.}~\bibnamefont{Zhang}},
  \bibinfo{author}{\bibfnamefont{I.-K.} \bibnamefont{Sou}},
  \bibinfo{author}{\bibfnamefont{G.~K.~L.} \bibnamefont{Wong}},
  \bibinfo{author}{\bibfnamefont{J.-N.} \bibnamefont{Wang}},
  \bibinfo{author}{\bibfnamefont{H.-Z.} \bibnamefont{Lu}},
  \bibinfo{author}{\bibfnamefont{S.-Q.} \bibnamefont{Shen}}, \bibnamefont{and}
  \bibinfo{author}{\bibfnamefont{F.-C.} \bibnamefont{Zhang}},
  \bibinfo{journal}{Physical Review Letters} \textbf{\bibinfo{volume}{106}},
  \bibinfo{pages}{166805} (\bibinfo{year}{2011}).

\bibitem[{\citenamefont{Zhao et~al.}(2013)\citenamefont{Zhao, Chang, Jiang,
  DaSilva, Sun, Wang, Xing, Wang, He, Ma et~al.}}]{Zhao13}
\bibinfo{author}{\bibfnamefont{Y.}~\bibnamefont{Zhao}},
  \bibinfo{author}{\bibfnamefont{C.-Z.} \bibnamefont{Chang}},
  \bibinfo{author}{\bibfnamefont{Y.}~\bibnamefont{Jiang}},
  \bibinfo{author}{\bibfnamefont{A.}~\bibnamefont{DaSilva}},
  \bibinfo{author}{\bibfnamefont{Y.}~\bibnamefont{Sun}},
  \bibinfo{author}{\bibfnamefont{H.}~\bibnamefont{Wang}},
  \bibinfo{author}{\bibfnamefont{Y.}~\bibnamefont{Xing}},
  \bibinfo{author}{\bibfnamefont{Y.}~\bibnamefont{Wang}},
  \bibinfo{author}{\bibfnamefont{K.}~\bibnamefont{He}},
  \bibinfo{author}{\bibfnamefont{X.}~\bibnamefont{Ma}}, \bibnamefont{et~al.},
  \bibinfo{journal}{Scientific Reports} \textbf{\bibinfo{volume}{3}},
  \bibinfo{pages}{3060} (\bibinfo{year}{2013}).

\bibitem[{\citenamefont{Lin et~al.}(2013)\citenamefont{Lin, He, Liao, Wang,
  Sacksteder~IV, Yang, Guan, Zhang, Gu, Zhang et~al.}}]{Lin13}
\bibinfo{author}{\bibfnamefont{C.}~\bibnamefont{Lin}},
  \bibinfo{author}{\bibfnamefont{X.}~\bibnamefont{He}},
  \bibinfo{author}{\bibfnamefont{J.}~\bibnamefont{Liao}},
  \bibinfo{author}{\bibfnamefont{X.}~\bibnamefont{Wang}},
  \bibinfo{author}{\bibfnamefont{V.}~\bibnamefont{Sacksteder~IV}},
  \bibinfo{author}{\bibfnamefont{W.}~\bibnamefont{Yang}},
  \bibinfo{author}{\bibfnamefont{T.}~\bibnamefont{Guan}},
  \bibinfo{author}{\bibfnamefont{Q.}~\bibnamefont{Zhang}},
  \bibinfo{author}{\bibfnamefont{L.}~\bibnamefont{Gu}},
  \bibinfo{author}{\bibfnamefont{G.}~\bibnamefont{Zhang}},
  \bibnamefont{et~al.}, \bibinfo{journal}{Physical Review B}
  \textbf{\bibinfo{volume}{88}}, \bibinfo{pages}{041307}
  (\bibinfo{year}{2013}).

\bibitem[{\citenamefont{Wang et~al.}(2012)\citenamefont{Wang, He, Guan, Liao,
  Lin, Wu, Li, and Zeng}}]{Wang12}
\bibinfo{author}{\bibfnamefont{X.}~\bibnamefont{Wang}},
  \bibinfo{author}{\bibfnamefont{X.}~\bibnamefont{He}},
  \bibinfo{author}{\bibfnamefont{T.}~\bibnamefont{Guan}},
  \bibinfo{author}{\bibfnamefont{J.}~\bibnamefont{Liao}},
  \bibinfo{author}{\bibfnamefont{C.}~\bibnamefont{Lin}},
  \bibinfo{author}{\bibfnamefont{K.}~\bibnamefont{Wu}},
  \bibinfo{author}{\bibfnamefont{Y.}~\bibnamefont{Li}}, \bibnamefont{and}
  \bibinfo{author}{\bibfnamefont{C.}~\bibnamefont{Zeng}},
  \bibinfo{journal}{Physica E: Low-dimensional Systems and Nanostructures}
  \textbf{\bibinfo{volume}{46}}, \bibinfo{pages}{236} (\bibinfo{year}{2012}).

\bibitem[{\citenamefont{Lee et~al.}(2012)\citenamefont{Lee, Park, Lee, Kim, and
  Lee}}]{Lee12}
\bibinfo{author}{\bibfnamefont{J.}~\bibnamefont{Lee}},
  \bibinfo{author}{\bibfnamefont{J.}~\bibnamefont{Park}},
  \bibinfo{author}{\bibfnamefont{J.-H.} \bibnamefont{Lee}},
  \bibinfo{author}{\bibfnamefont{J.~S.} \bibnamefont{Kim}}, \bibnamefont{and}
  \bibinfo{author}{\bibfnamefont{H.-J.} \bibnamefont{Lee}},
  \bibinfo{journal}{Physical Review B} \textbf{\bibinfo{volume}{86}},
  \bibinfo{pages}{245321} (\bibinfo{year}{2012}).

\bibitem[{\citenamefont{Cha et~al.}(2012)\citenamefont{Cha, Kong, Hong,
  Analytis, Lai, and Cui}}]{Cha12}
\bibinfo{author}{\bibfnamefont{J.~J.} \bibnamefont{Cha}},
  \bibinfo{author}{\bibfnamefont{D.}~\bibnamefont{Kong}},
  \bibinfo{author}{\bibfnamefont{S.-S.} \bibnamefont{Hong}},
  \bibinfo{author}{\bibfnamefont{J.~G.} \bibnamefont{Analytis}},
  \bibinfo{author}{\bibfnamefont{K.}~\bibnamefont{Lai}}, \bibnamefont{and}
  \bibinfo{author}{\bibfnamefont{Y.}~\bibnamefont{Cui}}, \bibinfo{journal}{Nano
  Letters} \textbf{\bibinfo{volume}{12}}, \bibinfo{pages}{1107}
  (\bibinfo{year}{2012}).

\bibitem[{\citenamefont{Gehring et~al.}(2013)\citenamefont{Gehring, Benia,
  Weng, Dinnebier, Ast, Burghard, and Kern}}]{Gehring13}
\bibinfo{author}{\bibfnamefont{P.}~\bibnamefont{Gehring}},
  \bibinfo{author}{\bibfnamefont{H.~M.} \bibnamefont{Benia}},
  \bibinfo{author}{\bibfnamefont{Y.}~\bibnamefont{Weng}},
  \bibinfo{author}{\bibfnamefont{R.}~\bibnamefont{Dinnebier}},
  \bibinfo{author}{\bibfnamefont{C.~R.} \bibnamefont{Ast}},
  \bibinfo{author}{\bibfnamefont{M.}~\bibnamefont{Burghard}}, \bibnamefont{and}
  \bibinfo{author}{\bibfnamefont{K.}~\bibnamefont{Kern}},
  \bibinfo{journal}{Nano Letters} \textbf{\bibinfo{volume}{13}},
  \bibinfo{pages}{1179} (\bibinfo{year}{2013}).

\bibitem[{\citenamefont{Chen et~al.}(2014)\citenamefont{Chen, Chen, Schouteden,
  Huang, Wang, Li, Miao, Wang, Li, Zhao et~al.}}]{Chen14}
\bibinfo{author}{\bibfnamefont{T.}~\bibnamefont{Chen}},
  \bibinfo{author}{\bibfnamefont{Q.}~\bibnamefont{Chen}},
  \bibinfo{author}{\bibfnamefont{K.}~\bibnamefont{Schouteden}},
  \bibinfo{author}{\bibfnamefont{W.}~\bibnamefont{Huang}},
  \bibinfo{author}{\bibfnamefont{X.}~\bibnamefont{Wang}},
  \bibinfo{author}{\bibfnamefont{Z.}~\bibnamefont{Li}},
  \bibinfo{author}{\bibfnamefont{F.}~\bibnamefont{Miao}},
  \bibinfo{author}{\bibfnamefont{X.}~\bibnamefont{Wang}},
  \bibinfo{author}{\bibfnamefont{Z.}~\bibnamefont{Li}},
  \bibinfo{author}{\bibfnamefont{B.}~\bibnamefont{Zhao}}, \bibnamefont{et~al.},
  \bibinfo{journal}{arXiv.org} \textbf{\bibinfo{volume}{arXiv:1401.0383v1}}
  (\bibinfo{year}{2014}).

\bibitem[{\citenamefont{Sacksteder et~al.}(2012)\citenamefont{Sacksteder,
  Kettemann, Wu, Dai, and Fang}}]{sacksteder2012spin}
\bibinfo{author}{\bibfnamefont{V.~E.} \bibnamefont{Sacksteder}},
  \bibinfo{author}{\bibfnamefont{S.}~\bibnamefont{Kettemann}},
  \bibinfo{author}{\bibfnamefont{Q.~S.} \bibnamefont{Wu}},
  \bibinfo{author}{\bibfnamefont{X.}~\bibnamefont{Dai}}, \bibnamefont{and}
  \bibinfo{author}{\bibfnamefont{Z.}~\bibnamefont{Fang}},
  \bibinfo{journal}{Physical Review B} \textbf{\bibinfo{volume}{85}},
  \bibinfo{pages}{205303} (\bibinfo{year}{2012}).

\bibitem[{\citenamefont{Bergmann}(1989)}]{bergmann1989weak}
\bibinfo{author}{\bibfnamefont{G.}~\bibnamefont{Bergmann}},
  \bibinfo{journal}{Physical Review B} \textbf{\bibinfo{volume}{39}},
  \bibinfo{pages}{11280} (\bibinfo{year}{1989}).

\bibitem[{\citenamefont{Sacksteder and Dai}()}]{MyInplane}
\bibinfo{author}{\bibfnamefont{V.}~\bibnamefont{Sacksteder}} \bibnamefont{and}
  \bibinfo{author}{\bibfnamefont{X.}~\bibnamefont{Dai}}, \bibinfo{note}{in
  preparation}.

\bibitem[{\citenamefont{Maekawa and
  Fukuyama}(1981)}]{maekawa1981magnetoresistance}
\bibinfo{author}{\bibfnamefont{S.}~\bibnamefont{Maekawa}} \bibnamefont{and}
  \bibinfo{author}{\bibfnamefont{H.}~\bibnamefont{Fukuyama}},
  \bibinfo{journal}{Journal of the Physical Society of Japan}
  \textbf{\bibinfo{volume}{50}}, \bibinfo{pages}{2516} (\bibinfo{year}{1981}).

\bibitem[{\citenamefont{Lee and Ramakrishnan}(1985)}]{lee1985disordered}
\bibinfo{author}{\bibfnamefont{P.~A.} \bibnamefont{Lee}} \bibnamefont{and}
  \bibinfo{author}{\bibfnamefont{T.}~\bibnamefont{Ramakrishnan}},
  \bibinfo{journal}{Reviews of Modern Physics} \textbf{\bibinfo{volume}{57}},
  \bibinfo{pages}{287} (\bibinfo{year}{1985}).

\bibitem[{\citenamefont{Altshuler et~al.}(1982)\citenamefont{Altshuler, Aronov,
  and Zuzin}}]{altshuler1982spin}
\bibinfo{author}{\bibfnamefont{B.}~\bibnamefont{Altshuler}},
  \bibinfo{author}{\bibfnamefont{A.}~\bibnamefont{Aronov}}, \bibnamefont{and}
  \bibinfo{author}{\bibfnamefont{A.~Y.} \bibnamefont{Zuzin}},
  \bibinfo{journal}{Solid State Communications} \textbf{\bibinfo{volume}{44}},
  \bibinfo{pages}{137} (\bibinfo{year}{1982}).

\bibitem[{\citenamefont{Sahnoune et~al.}(1992)\citenamefont{Sahnoune,
  Str{\"o}m-Olsen, and Fischer}}]{sahnoune1992influence}
\bibinfo{author}{\bibfnamefont{A.}~\bibnamefont{Sahnoune}},
  \bibinfo{author}{\bibfnamefont{J.~O.} \bibnamefont{Str{\"o}m-Olsen}},
  \bibnamefont{and} \bibinfo{author}{\bibfnamefont{H.~E.}
  \bibnamefont{Fischer}}, \bibinfo{journal}{Physical Review B}
  \textbf{\bibinfo{volume}{46}}, \bibinfo{pages}{10035} (\bibinfo{year}{1992}).

\bibitem[{\citenamefont{Rammer}(2004)}]{rammer2004quantum}
\bibinfo{author}{\bibfnamefont{J.}~\bibnamefont{Rammer}},
  \emph{\bibinfo{title}{Quantum transport theory}}
  (\bibinfo{publisher}{Westview Press}, \bibinfo{year}{2004}).

\bibitem[{\citenamefont{McCann et~al.}(2006)\citenamefont{McCann, Kechedzhi,
  FalÕko, Suzuura, Ando, and Altshuler}}]{mccann2006weak}
\bibinfo{author}{\bibfnamefont{E.}~\bibnamefont{McCann}},
  \bibinfo{author}{\bibfnamefont{K.}~\bibnamefont{Kechedzhi}},
  \bibinfo{author}{\bibfnamefont{V.~I.} \bibnamefont{FalÕko}},
  \bibinfo{author}{\bibfnamefont{H.}~\bibnamefont{Suzuura}},
  \bibinfo{author}{\bibfnamefont{T.}~\bibnamefont{Ando}}, \bibnamefont{and}
  \bibinfo{author}{\bibfnamefont{B.~L.} \bibnamefont{Altshuler}},
  \bibinfo{journal}{Physical review letters} \textbf{\bibinfo{volume}{97}},
  \bibinfo{pages}{146805} (\bibinfo{year}{2006}).

\bibitem[{\citenamefont{Zhang et~al.}(2012)\citenamefont{Zhang, Kane, and
  Mele}}]{zhang2012surface}
\bibinfo{author}{\bibfnamefont{F.}~\bibnamefont{Zhang}},
  \bibinfo{author}{\bibfnamefont{C.~L.} \bibnamefont{Kane}}, \bibnamefont{and}
  \bibinfo{author}{\bibfnamefont{E.~J.} \bibnamefont{Mele}},
  \bibinfo{journal}{Physical Review B} \textbf{\bibinfo{volume}{86}},
  \bibinfo{pages}{081303} (\bibinfo{year}{2012}).

\bibitem[{\citenamefont{Silvestrov et~al.}(2012)\citenamefont{Silvestrov,
  Brouwer, and Mishchenko}}]{silvestrov2012spin}
\bibinfo{author}{\bibfnamefont{P.~G.} \bibnamefont{Silvestrov}},
  \bibinfo{author}{\bibfnamefont{P.~W.} \bibnamefont{Brouwer}},
  \bibnamefont{and} \bibinfo{author}{\bibfnamefont{E.~G.}
  \bibnamefont{Mishchenko}}, \bibinfo{journal}{Physical Review B}
  \textbf{\bibinfo{volume}{86}}, \bibinfo{pages}{075302}
  (\bibinfo{year}{2012}).

\bibitem[{\citenamefont{Nordheim}(1931)}]{VCA1}
\bibinfo{author}{\bibfnamefont{L.}~\bibnamefont{Nordheim}},
  \bibinfo{journal}{Ann. Phys. (Leipzig)} \textbf{\bibinfo{volume}{9}},
  \bibinfo{pages}{607} (\bibinfo{year}{1931}).

\bibitem[{\citenamefont{Muto}(1938)}]{VCA2}
\bibinfo{author}{\bibfnamefont{T.}~\bibnamefont{Muto}}, \bibinfo{journal}{Sci.
  Pap. Inst. Phys. Chem. Res. (Jpn.)} \textbf{\bibinfo{volume}{34}},
  \bibinfo{pages}{377} (\bibinfo{year}{1938}).

\bibitem[{\citenamefont{Kettemann and
  Mazzarello}(2002)}]{kettemann2002magnetolocalization}
\bibinfo{author}{\bibfnamefont{S.}~\bibnamefont{Kettemann}} \bibnamefont{and}
  \bibinfo{author}{\bibfnamefont{R.}~\bibnamefont{Mazzarello}},
  \bibinfo{journal}{Physical Review B} \textbf{\bibinfo{volume}{65}},
  \bibinfo{pages}{085318} (\bibinfo{year}{2002}).

\bibitem[{\citenamefont{Kettemann}(2007)}]{kettemann2007dimensional}
\bibinfo{author}{\bibfnamefont{S.}~\bibnamefont{Kettemann}},
  \bibinfo{journal}{Physical review letters} \textbf{\bibinfo{volume}{98}},
  \bibinfo{pages}{176808} (\bibinfo{year}{2007}).

\bibitem[{\citenamefont{Sharvin}(1984)}]{Sharvin84}
\bibinfo{author}{\bibfnamefont{Y.~V.} \bibnamefont{Sharvin}},
  \bibinfo{journal}{Physica B} \textbf{\bibinfo{volume}{126}},
  \bibinfo{pages}{288} (\bibinfo{year}{1984}).

\end{thebibliography}

\end{document}